\documentclass{egpubl}
\usepackage{eg2025}
 
\ConferencePaper        
\CGFStandardLicense

\usepackage[T1]{fontenc}
\usepackage{dfadobe}  
\usepackage{cite}  
\BibtexOrBiblatex
\electronicVersion
\PrintedOrElectronic
\ifpdf \usepackage[pdftex]{graphicx} \pdfcompresslevel=9
\else \usepackage[dvips]{graphicx} \fi

\usepackage{egweblnk} 

\usepackage{soul} 
\usepackage{xcolor} 
\usepackage[labelfont=bf,textfont=it]{caption}
\usepackage[labelfont=bf,textfont=it]{subcaption}
\usepackage{amsfonts, amsmath, amssymb}
\usepackage{booktabs}
\usepackage{makecell}

\usepackage{array}
\usepackage{multirow}
\usepackage{url}

\title[ASMR: Adaptive Skeleton-Mesh Rigging and Skinning via 2D Generative Prior]%
      {ASMR: Adaptive Skeleton-Mesh Rigging and Skinning \\via 2D Generative Prior}
\author[Hong et al.]{\parbox{\textwidth}{\centering
        Seokhyeon Hong$^{*}$\orcid{0000-0002-8490-5338} \quad\quad
        Soojin Choi$^{*}$\orcid{0000-0002-0832-6545}  \quad\quad 
        Chaelin Kim\orcid{0000-0001-8355-7522}  \quad\quad 
        Sihun Cha\orcid{0000-0001-9506-9438}  \quad\quad 
        Junyong Noh\orcid{0000-0003-1925-3326}\\
        \vspace{0.3cm}
    {\parbox{\textwidth}{\centering Visual Media Lab, KAIST}
}
}}

\begin{document}
\maketitle
\begin{abstract}
Despite the growing accessibility of skeletal motion data, integrating it for animating character meshes remains challenging due to diverse configurations of both skeletons and meshes. Specifically, the body scale and bone lengths of the skeleton should be adjusted in accordance with the size and proportions of the mesh, ensuring that all joints are accurately positioned within the character mesh. Furthermore, defining skinning weights is complicated by variations in skeletal configurations, such as the number of joints and their hierarchy, as well as differences in mesh configurations, including their connectivity and shapes. While existing approaches have made efforts to automate this process, they hardly address the variations in both skeletal and mesh configurations. In this paper, we present a novel method for the automatic rigging and skinning of character meshes using skeletal motion data, accommodating arbitrary configurations of both meshes and skeletons. The proposed method predicts the optimal skeleton aligned with the size and proportion of the mesh as well as defines skinning weights for various mesh-skeleton configurations, without requiring explicit supervision tailored to each of them. By incorporating Diffusion 3D Features (Diff3F) as semantic descriptors of character meshes, our method achieves robust generalization across different configurations. To assess the performance of our method in comparison to existing approaches, we conducted comprehensive evaluations encompassing both quantitative and qualitative analyses, specifically examining the predicted skeletons, skinning weights, and deformation quality.
\end{abstract}  

\def\thefootnote{*}\footnotetext{Equal contribution}\def\thefootnote{\arabic{footnote}}
\section{Introduction}

Skeletal data has long been a fundamental representation for character animation. While capturing the skeletal animation was traditionally challenging due to its reliance on costly methods like manual keyframing and motion capture, recent advancements have simplified this process. For example, high-fidelity human motion can now be captured with a minimal setup using video-based 3D pose estimation~\cite{pavllo20193d, zheng20213d, shin2024wham, rempe2021humor} or sensor-based motion capture systems~\cite{yi2024physical, ponton2023sparseposer, huang2018deep}. Additionally, the growing availability of large public motion capture datasets~\cite{CMU, komura2017recurrent, starke2019neural, harvey2020robust, mason2022local} has made high-quality motion data easily accessible, significantly reducing production costs.

Despite the increased accessibility of skeletal motion data, integrating it to various character meshes remains a significant challenge. To animate a character mesh using skeletal data, the mesh must first undergo a process called rigging, which involves aligning the skeleton with the mesh and defining skinning weights that determine how each vertex of the mesh is influenced by the movement of the underlying joints. Because the motion capture data typically has varying body scales and bone lengths, it is necessary to adjust the skeleton to match the proportions of the mesh, followed by retargeting the original sequence to the adjusted skeleton. Otherwise, unexpected deformation can be made due to joints not being aligned with the mesh, as exemplified by a case when a shoulder joint is located outside the mesh or embedded inside the torso. Furthermore, skeletons and meshes each possesses distinct configurations: skeletons vary in terms of bone length, the number of joints and their hierarchical connectivity, whereas meshes exhibit different vertex connectivity and body volume. 
These discrepancies and the inherent inhomogeneity between skeleton and mesh data necessitate a careful designation of skinning weights tailored to the specific configurations of both the skeleton and mesh. This complexity underscores the need for an approach that can automatically animate character meshes using existing skeletal data. For clarification, we define \textbf{skeleton configuration} as encompassing (i) \textbf{shape}, referring to geometric properties such as body scale and bone lengths and (ii) \textbf{structure}, which includes the number of joints and their hierarchy. Similarly, \textbf{mesh configuration} is defined as (i) \textbf{shape}, representing the body proportions and volume and (ii) \textbf{structure}, denoting the vertex connectivity.

\begin{figure}[t]
\centering
\includegraphics[width=\columnwidth]{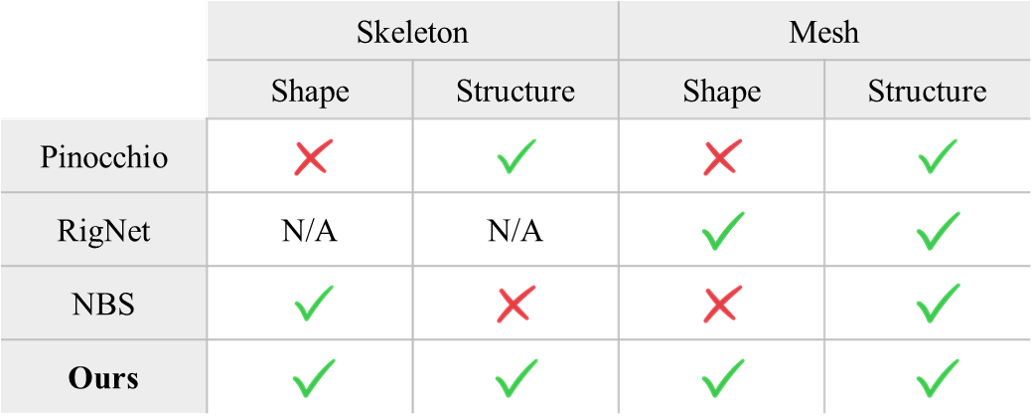}
\caption{Comparison of the robustness of different auto-rigging and skinning approaches to variations in input configurations.
While each approach has limited robustness or is not applicable in at least one component, the proposed method achieves robustness across all components.}
\label{fig:key_problem}
\vspace{-1.5em}
\end{figure}

To address these challenges, auto-rigging methods have been developed to automatically estimate the skeleton and skinning weights that are aligned to a given character mesh. For instance, some approaches predict skinning weights assuming that a skeleton matching the mesh in size and proportions is already provided~\cite{mosella2022skinningnet, wang2024towards}, while others allow the skeleton to differ in size and proportion from the mesh but limit the number of joints and their connectivity to a pre-defined template~\cite{li2021nbs}. Yet other approaches derive both the skeleton and skinning weights directly from the mesh, allowing skeletons with varying numbers of joints and connectivity, though with restricted flexibility as they do not allow explicit specification of the exact number of joints or their hierarchical structures~\cite{xu2020rignet}. Pinocchio~\cite{baran2007automatic} facilitates the use of desired skeletons with varying numbers of joints for animating the character mesh; however, this approach assumes that the input mesh has proportions approximately similar to those of the given skeleton. While the aforementioned approaches have demonstrated the capability of producing promising results, they all fall short in simultaneously addressing variations in both skeletal and mesh configurations, as summarized in Figure ~\ref{fig:key_problem}.

In this work, we propose a novel method for automatic rigging and skinning of character meshes to animate them using skeletal motion data. Our approach is designed to accommodate arbitrary configurations for both the mesh and skeleton, operating through two key modules: \textbf{Skeletal Articulation Prediction} and \textbf{Skinning Weight Prediction}. The Skeletal Articulation Prediction module first identifies the optimal target skeletal articulation, ensuring that it has the same number of joints and hierarchy as the source skeleton. By employing an attention mechanism to capture the relationships between the input mesh and skeleton, this module is trained to predict a target skeleton that aligns with the mesh while preserving the original joint connectivity. Subsequently, the source skeletal motion is retargeted to the target skeleton, generating the motion that drives the movement of the character mesh. To address the challenge of predicting skinning weights for diverse mesh and skeleton configurations, we introduce a Skinning Weight Prediction module.
This module implicitly learns skinning weights from deformed meshes, avoiding the necessity for explicit ground truth skinning weights tailored to each specific mesh and skeleton configuration. Finally, using the predicted skinning weights, the character mesh is animated along with the target skeletal data through Linear Blend Skinning (LBS).

The acquisition of extensive 3D character datasets that adequately encompass the full range of mesh variations presents a significant challenge. Additionally, geometric features of character meshes, such as vertex positions and normals, do not fully capture the semantic information of each vertex, as the same vertex position of different characters could correspond to different body parts depending on their proportions. To address this limitation, we leverage Diffusion 3D Features (Diff3F) ~\cite{luo2024diffusion}, which are semantic descriptors that are derived from foundational vision models pre-trained on large-scale datasets. By incorporating the diffusion-based features that establish consistent semantic correspondences across diverse input meshes, our method effectively associates vertices with skeletal joints, even in unseen mesh configurations. This enables our method to achieve robust generalization across a wide range of character meshes.

Our technical contributions can be summarized as follows:
\begin{itemize}
  \item We propose attention-based Skeletal Articulation Prediction and Skinning Weight Prediction modules that capture the inter-relationships between an arbitrary number of skeleton joints and mesh vertices, enabling rigging and skinning character meshes from skeletons having arbitrary configurations.
  \item We present a self-supervised learning approach that facilitates the implicit learning of skinning weights for diverse mesh-skeleton configurations, eliminating the need for the ground truth skinning weights tailored to each configurations.
  \item We propose a novel method that leverages an image generative feature as a semantic prior for the task of auto-rigging and skinning character meshes with skeletal data, enhancing the ability to generalize to unseen 3D character meshes.
\end{itemize}

\section{Related Work}
\subsection{Mesh Deformation with Skeletal Articulations}
In computer graphics, deforming a mesh according to a given skeletal animation has long been a significant challenge. Various skinning-based approaches have been developed to tackle this problem. LBS~\cite{magnenat1988joint} is one of the most widely used techniques due to its simplicity and computational efficiency. Dual Quaternion Skinning (DQS)~\cite{hejl2004hardware, kavan2007skinning, le2014robust} better preserves rotational properties, leading to smoother deformations. Spherical-based skinning~\cite{kavan2005spherical} further extends these concepts by leveraging geometric properties to enhance deformation quality. Multi-linear techniques~\cite{wang2002multi, merry2006animation} address non-linear deformations with improved accuracy and effectively reduce artifacts like volume loss. To animate a character mesh using these skinning approaches, the mesh must first be rigged to have an aligned skeleton and corresponding skinning weights for each vertex. 

To alleviate the burden involved in rigging and skinning, which often requires expertise, auto-rigging techniques have been proposed. For human-like biped characters, 3D animation applications provide auto-rigging tools that fit a template skeleton to a target character mesh~\cite{autodesk2018quick, mixamo}. These tools often require users to manually specify the positions of joints or key points, which can be cumbersome. Pinocchio~\cite{baran2007automatic} is the first research to automate the entire rigging and skinning process. It first performs fitting a user-specified skeleton to a character mesh by iteratively contracting the mesh until it converges to a skeleton-like graph, and then calculates skinning weights for each vertex using a heat diffusion model. Subsequent research has expanded to handle character models in a part-wise manner~\cite{miller2010frankenrigs,seo2010rigging} or to accommodate more complex character models with varying proportions or morphologies~\cite{bharaj2012automatically}. FAKIR~\cite{fu2020fakir} proposed an iterative algorithm for skeleton registration that reveals the anatomy and pose given raw points, including scanned statues of clothed humans, animals, or hybrid figures. Upon these foundational studies, Neural Blend Shape (NBS)~\cite{li2021nbs} was introduced as a data-driven deep learning approach. It utilizes a neural network to generate rigged and posed meshes by predicting joint offsets of the template skeleton and pose-dependent blendshapes to enhance the quality of the animated mesh. While NBS offers a flexible and adaptive rigging solution, its effectiveness is constrained by its reliance on a pre-defined skeleton structure and meshes that are closely aligned with the SMPL distribution. Estimating additional bones~\cite{ma2023tarig} and separating the skinning and deformation of clothing from the body~\cite{wang2024towards} improved the mesh deformation quality for highly complex character meshes with apparels and accessories. These methods are designed to work with a fixed template skeleton.

In contrast to approaches that require a specific target template skeleton, some research has focused on estimating the skeleton directly from the geometry data itself. Early methods in this domain often relied on geometric techniques, such as Laplacian contraction methods, to derive a skeleton~\cite{au2008skeleton, cao2010point}. RigMesh~\cite{borosan2012rigmesh} segments the model, generates a skeleton for each segment, and integrates the segmented components into a cohesive structure. More recently, advances in deep learning have enabled predicting skeleton directly from 3D geometry using deep neural networks. RigNet~\cite{xu2020rignet}, which employs Graph Neural Network~(GNN), provides a robust solution for complex and diverse character models. MoRig~\cite{xu2022morig} built upon this by incorporating motion features extracted from point cloud animations, enhancing skeleton estimation performance and enabling effective animation retargeting. While these approaches focused on generating random skeletal structures, our method leverages skeleton structures as input, to provide users with control over the generated structure. For this, we utilize attention maps to capture the inter-relationship between vertices and joints, allowing for the flexibility of handling arbitrary numbers of mesh vertices and skeleton joints. Additionally, we leverage generative features obtained from large-scale image models as semantic priors for each vertex, incorporating this semantic information when estimating skinning weights and joint offsets.

\subsection{Mesh Deformation Transfer}
Another branch of research focuses on animating bipedal character meshes without relying on skeletons, instead directly transferring the deformation from the source to the target mesh by assuming consistent mesh connectivity between the source and target meshes~\cite{groueix20183d, tan2018variational, gao2018automatic}. More advanced approaches enable retargeting between character meshes with different connectivities by utilizing local deformations, such as per-triangle Jacobian~\cite{aigerman2022njf}, implicit skinning weights~\cite{liao2022sfpt, wang2023hmc}, and per-vertex displacement~\cite{wang2023zpt}. While these studies enable the animating and retargeting of unrigged character meshes, the learned latent spaces that represent character poses and deformation are not interpretable, making it difficult to manipulate or adjust the deformed results. To address this, we aim to leverage the intuitive and editable nature of skeleton representations and skinning weights. By estimating the skeleton and skinning weights for the given mesh, we enable both intuitive editing and effective pose transfer.

\subsection{Skeleton-based Motion Retargeting}
To reuse existing skeletal motion data for various characters, the motion retargeting technique, which involves adapting the original sequence to fit different skeletons, has been extensively studied. Early work focused on retargeting motions to target skeletons having different bone lengths and proportions through non-linear optimization with a set of constraints to preserve essential characteristics involved in the source motion~\cite{gleicher1998retarget, choi-ko2000online, lee1999hierarchical, shin2001computer}. With the increased availability of skeletal motion data, deep learning-based methods have addressed this with a data-driven approach, making desired joint transformations to be predicted by neural networks~\cite{villegas2018neural, lim2019pmnet, villegas2021contact, zhang2023skinned}. Because skeletons can differ not only in body proportions but also in the number of joints and their hierarchy, more advanced neural network architectures have also been proposed to handle a wider range of skeletal configurations. By embedding skeletal motion from different skeletons into a shared latent space, Skeleton-aware networks~(SAN)~\cite{aberman2020skeleton} enabled motion transfer between skeletons that are topologically equivalent but with different numbers of joints. While this work requires a separate network to be trained for each skeletal configuration, SAME~\cite{lee2023same} constructed a skeleton-agnostic motion embedding using an autoencoder based on graph convolution networks (GCN), enabling arbitrary skeletons to be processed within a single network. While these studies effectively predict joint transformations for target skeletons to best preserve the source skeletal motion, animating the character meshes with the retargeted results requires aligning the target skeleton to the mesh and predicting corresponding skinning weights. To address this problem, we propose an end-to-end framework that generates proper skeletal articulation and skinning weights of the input mesh simultaneously. Once the articulation is predicted, we transfer the source skeletal motion data to the predicted articulation using a pre-trained SAME~\cite{lee2023same}. Subsequently, we apply the retargeted skeletal motion to the character mesh through the predicted skinning weights, allowing the character mesh to be directly animated by the given skeletal motion data. 
\section{Method}
\begin{figure*}[t]
    \centering    \includegraphics[width=\textwidth]{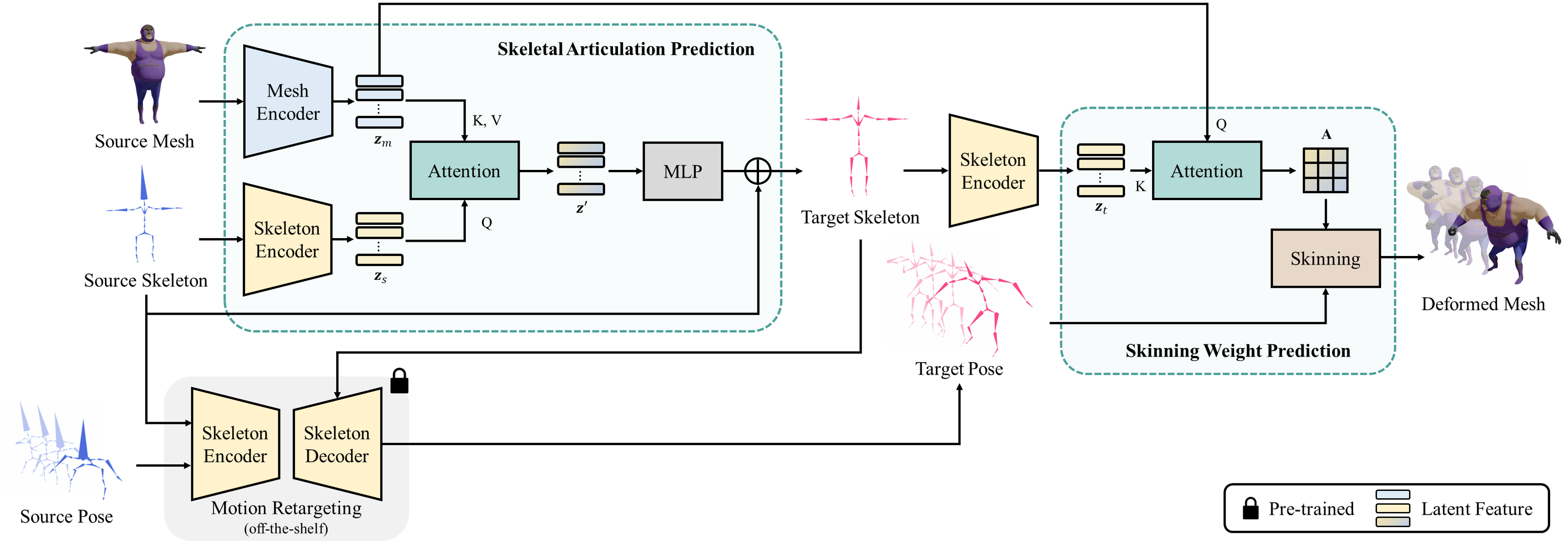}
    \caption{Overview of our method. Given a source mesh to be animated and a source skeletal motion to derive its movement, our method predicts the target skeleton and the corresponding skinning weights that generate plausible deformation of the character mesh in accordance with the source skeletal movement. To accommodate source skeletons with arbitrary structures, we leverage an off-the-shelf retargeting module that aligns the target skeleton to the source pose, generating the target pose. Finally, the target pose, combined with the predicted skinning weights, is used to deform the character mesh. While our method does not rely on textural information, textures on the character meshes are included only to illustrate different poses.}
    \label{fig:overview}
    \vspace{-1.5em}
\end{figure*}

Our goal is to animate a bipedal mesh using available skeletal motion data, with a focus on flexibility and generalization. To achieve this, we accommodate variations in skeletal configurations including differences in body scale, bone lengths, the number of joints, and their hierarchies as well as meshes with varying connectivity and shapes.  As shown in Figure~\ref{fig:overview}, our approach consists of two main components: \textbf{Skeletal Articulation Prediction} and \textbf{Skinning Weight Prediction}. The Skeletal Articulation Prediction module predicts a target skeleton that aligns with the input mesh in terms of the sizes and body proportions while preserving the number of joints and their hierarchy of the input skeleton. Note that the source mesh and source skeleton are consistently provided in a T-pose as a reference for calculating the posed state to ensure proper alignment for mesh deformation using LBS across all samples. In this stage, we leverage a 2D generative prior to obtain 3D deep features given the input mesh to improve the understanding of the model in terms of the semantic alignment between mesh and skeleton, resulting in generalizability to unseen shapes of character mesh. Subsequently, the Skinning Weight Prediction module estimates the skinning weights of the input mesh given the target skeleton using an attention mechanism. Finally, we deform the character mesh by applying LBS based on these predicted skinning weights. In the following sections, we outline the data representation used in our approach~(Section~\ref{sec:data}), followed by the architecture of Skeletal Articulation Prediction~(Section~\ref{sec:skeletal}), and Skinning Weight Prediction~(Section~\ref{sec:skinning}), and finally, the training process with data preparation~(Section~\ref{sec:training}).

\subsection{Data Representation}
\label{sec:data}
This section outlines the input data representation used in our framework, which remains consistent across both the training and inference phases. For each character model, we construct a skeletal motion data $M=({S, D^{1:N_T}})$, where $S$ denotes the skeleton, $D^{1:N_T}$ represents the motion data, and $N_T$ is the number of frames. Following Lee et al.~\cite{lee2023same}, $S$ is represented as follows:
\begin{align}
    S=\{ \mathbf{g}_{1:N_J}, \mathbf{o}_{1:N_J} \}, 
\end{align}
where $N_J$ is the number of joints. Each skeleton consists of $\mathbf{g}_{1:N_J}$ and $\mathbf{o}_{1:N_J}$, where $\mathbf{g}_j \in \mathbb{R}^{3}$ and $\mathbf{o}_j \in \mathbb{R}^3$ denote the global joint position at the rest pose and local joint position relative to its parent joint coordinate system, respectively. $D^t$ includes skeletal dynamics information at frame $t$. For more details on its elements and their derivation, please refer to Appendix~\ref{Appendix_A}.

The character mesh data $G$ is represented as follows:
\begin{equation}
    G= \{ \mathbf{V}_r, \mathbf{V}_d^{1:N_T}, \mathbf{V}_f \}.
\end{equation}
Here, $\mathbf{V}_r \in \mathbb{R}^{N_V \times 3}$ represents the vertex positions of the character mesh in the rest pose, where $N_V$ denotes the number of vertices. We then compute the deformed vertex positions $\mathbf{V}_d^{1:N_T}$ over timestep 1 to $N_T$ where $\mathbf{V}_d^t \in \mathbb{R}^{N_V \times 3}$ is obtained by deforming $\mathbf{V}_r$ using $D^{t}$ through the LBS formulation. Because $D^{t}$ is expressed with respect to the facing frame of the character, both $\mathbf{V}_r$ and $\mathbf{V}_d^t$ are expressed relative to the facing frame of the character mesh as well.

To obtain 3D diffusion features $\mathbf{V}_f$, we follow the pipeline of Diff3F~\cite{dutt2024diffusion}. Specifically, the character mesh is rendered using multi-view cameras, and we extract the diffusion features~\cite{tang2023emergent} and DINO features~\cite{oquab2023dinov2}. These features are fused and projected back onto the surface of the character mesh by aggregating the features of multiple views into one for each vertex, resulting in a 3D deep feature representation. By leveraging the large-scale image foundational models, the learned implicit features carry dense and accurate semantic priors~\cite{zhang2024tale, luo2024diffusion, hedlin2024unsupervised}. These high-level representations at each vertex contribute to significant improvements in both training efficiency and the overall performance of our method.

\begin{figure*}
    \centering    
    \includegraphics[width=\textwidth]{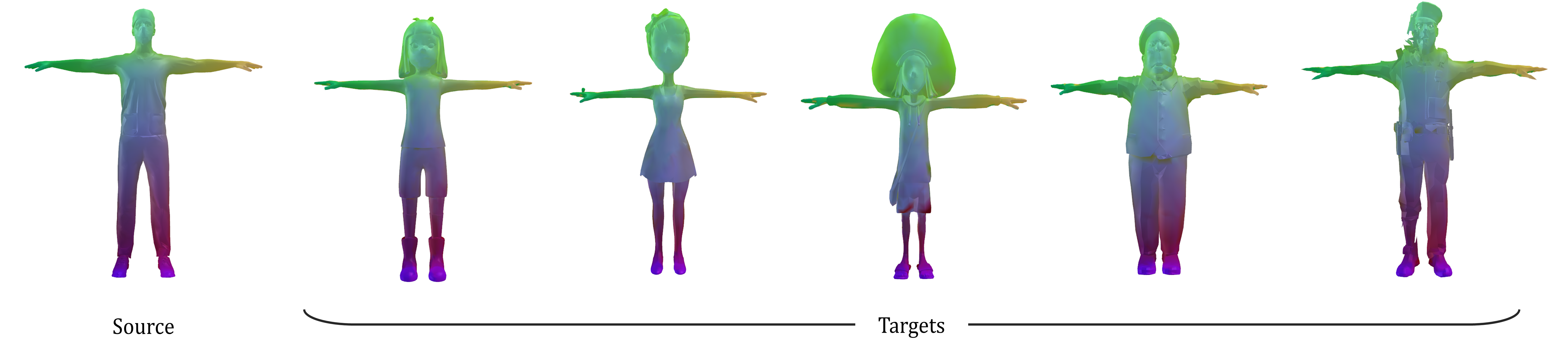}
    \caption{Visualizations of the vertex correspondences between characters using Diff3F, where corresponding points are similarly colored. The source character is shown on the left, and the target characters are on the right.}
    \label{fig:diff3f_corrspondence}
    \vspace{-1.5em}
\end{figure*}
\subsection{Skeletal Articulation Prediction}
\label{sec:skeletal}

\noindent\textbf{Mesh Encoder} \hspace{0.5mm}
The mesh encoder aims to estimate meaningful features from the input mesh, which effectively captures its correlation with a given skeleton, while being robust to variations in vertex connectivity and shapes. To achieve this, we employ the PointNet architecture~\cite{qi2017pointnet} for the mesh encoder, which processes the vertex positions at the rest pose $\mathbf{V}_{r}$ along with Diff3F $\mathbf{V}_{f}$, generating mesh latent features $\mathbf{z}_{m}\in \mathbb{R}^{N_V \times N_D}$, where $N_D$ denotes the dimension of the latent vector. Specifically, PointNet captures both local and global mesh features by combining information from individual vertices with that of the entire mesh, leading to a comprehensive understanding of the input mesh. Additionally, this architecture is flexible and generalizable to various vertex connectivities, even enabling the aggregation of features from unconnected components like the body and eyeballs.

To enhance the capability of the mesh encoder when capturing the semantic information of each vertex, we incorporate Diff3F~\cite{dutt2024diffusion} as an input, which provides accurate correspondences across 3D meshes with arbitrary connectivity by taking advantage of 2D generative prior. While incorporating a large-scale dataset of 3D character meshes could improve the capacity of the mesh encoder, acquiring such datasets is challenging. Instead, we leverage the generative priors of image diffusion models trained on large-scale image datasets, which embed rich semantic information, thereby circumventing the need for preparing such an extensive dataset. As shown in Figure~\ref{fig:diff3f_corrspondence}, Diff3F effectively captures accurate part-wise correspondences across different characters, even in the presence of highly stylized meshes. By combining both geometric and semantic features, our approach ensures that the encoder can adapt to diverse mesh configurations while accurately reflecting their relationships to the skeleton.

\noindent\textbf{Skeleton Encoder} \hspace{0.5mm}
We represent the skeleton as a graph, where the joints correspond to nodes, and the bones, which are connections between them, are represented as edges. In our framework, the body scale, bone length, the number of joints, and their hierarchies can be arbitrary, leading to inhomogeneous graph structures. To handle these variations, we employ GCN that learns via message passing by exchanging and propagating node features through the graph edges. Specifically, we employ graph attention networks~\cite{velivckovic2017graph} that leverage an attention mechanism to update the node features based on their neighbors. This approach enables us to derive a latent representation of the source skeleton, denoted as $\mathbf{z}_{s} \in \mathbb{R}^{N_J \times N_D}$, which maintains the original graph structure of the input skeleton while encoding its features and connectivity.

\noindent\textbf{Offset Residual Prediction} \hspace{0.5mm}
To ensure the generation of consistent animation when applying skeletal motion data to a character mesh, it is essential to align the size and body proportions of the mesh and those of the skeleton. Mismatches between these two can lead to distortions and artifacts, such as undesired stretching or compression of the animated mesh. This issue is especially critical in our framework because we allow arbitrary skeletons as input, which means that the skeleton and mesh may have different sizes and proportions.

To address this, we predict an offset residual that repositions the joints of the input skeleton by adjusting its local offset while maintaining its hierarchy. This results in a skeletal articulation aligned with the mesh, which we refer to as the target skeleton. Specifically, to capture the relationship between the skeleton and mesh data, we employ a cross-attention mechanism that can model the interactions between different modalities.
The cross-attention operation is defined as follows:
\begin{equation}
    \mathbf{z}' = \mathrm{softmax}\left( \frac{\mathbf{z}_{s} \mathbf{z}_{m}^{\top}}{\sqrt{N_D}} \right) \mathbf{z}_{m},
\end{equation}
where $\mathbf{z}_{s}$ and $\mathbf{z}_{m}$ represent the latent features of the source skeleton and mesh, respectively. The resulting latent feature $\mathbf{z}' \in \mathbb{R}^{N_J \times N_D}$ is then passed through a MLP to predict the residual of the local offset $\Delta\mathbf{o} \in \mathbb{R}^{N_J \times 3}$. To ensure the symmetry of the skeleton with respect to the lateral axis of the character, we update the residual of the local offset as follows:
\begin{equation}
    \Delta\mathbf{o}_{j} = \frac{1}{2}\left( \Delta\mathbf{o}_{j} + \Delta\mathbf{o}_{\rho(j)} \odot [-1, 1, 1] \right),
\end{equation}
where $\rho(j)$ denotes the index of the corresponding symmetrical joint of $j$-th joint and $\odot$ represents the element-wise multiplication.
For example, if $j$ refers to the left arm joint, $\rho(j)$ denotes the index of the corresponding right arm joint, while we define $\rho(j) = j$ for joints that do not have a symmetrical counterpart, such as spine joints.
The local offsets of the target skeleton, $\mathbf{o}_{tgt} \in \mathbb{R}^{N_J \times 3}$, are then computed as follows:
\begin{equation}
    \mathbf{o}_{tgt} = \mathbf{o}_{src} + \Delta\mathbf{o},
\end{equation}
where $\mathbf{o}_{src} \in \mathbb{R}^{N_J \times 3}$ represents the local offsets of the source skeleton.
Finally, to ensure that the skeleton is grounded, we translate the target skeleton along the up-axis so that the toe joints rest exactly on the ground.

\noindent\textbf{Motion Retargeting} \hspace{0.5mm}
Given a source pose, source skeleton, predicted target skeleton, and skinning weights of the source mesh, we aim to deform the source mesh according to the target pose, which follows the source pose with the target skeleton. To obtain the target pose $\hat{D}^t$, we employ a pre-trained SAME model~\cite{lee2023same}, which can accommodate heterogeneous input skeletons. While training a separate module for motion retargeting would be possible, it may necessitate complex simultaneous training of motion retargeting, rigging, and skinning. Because our primary focus is specifically on automating the rigging and skinning processes, we rely on an off-the-shelf method for motion retargeting to provide posed skeletons to deform the mesh. During training, we freeze the weights of the SAME model, allowing our network to focus exclusively on optimizing the target skeleton prediction and mesh deformation, while the motion retargeting is handled by the pre-trained capabilities of the SAME model.

\subsection{Skinning Weight Prediction}
\label{sec:skinning}
Skinning weights define how the movement of each joint influences the individual vertices of a mesh. Specifically, the deformation of each vertex is computed as a weighted sum of joint transformations relative to the rest pose. To ensure smooth and natural deformations, the skinning weights for each vertex must sum to one, resulting in a convex combination of joint transformations. To model this property within a neural network framework, we leverage the attention mechanism~\cite{vaswani2017attention}, whose principles align with skinning weights. The attention matrix, computed as the outer product of query and key vectors, encodes the relationships between components, which is analogous to how the skinning weight matrix captures the relationship between vertices and joints. Furthermore, the property that the relationships for each query component sum to one is also similar to the requirement that the skinning weights for each vertex sum to one.

We implement this by treating the latent feature of the mesh, $\mathbf{z}_{m}$, as the query, and the latent feature of the target skeleton, $\mathbf{z}_{t}\in\mathbb{R}^{N_J \times N_D}$, as the key. Note that $\mathbf{z}_{m}$ is the same latent vector used in Section~\ref{sec:skeletal}. The resulting attention matrix, $\mathbf{A} \in \mathbb{R}^{N_V \times N_J}$, is then computed as follows:
\begin{equation}
    \mathbf{A} = \mathrm{softmax}\left( \frac{\mathbf{z}_{m} \mathbf{z}_{t}^{\top}}{\sqrt{N_D}} \right).
\end{equation}
We interpret the learned attention weights as implicit skinning weights, which we apply within the LBS formulation to deform the character mesh.
Specifically, each vertex $\mathbf{V}_{r,i}$ is deformed as follows:
\begin{equation}
    \begin{bmatrix}
        \hat{\mathbf{V}}_{d, i} \\ 1
    \end{bmatrix} = \left( \sum_{j=1}^{N_J} \mathbf{A}_{i, j} \mathbf{T}_{j} \right) 
    \begin{bmatrix}
        \mathbf{V}_{r, i} \\ 1
    \end{bmatrix}
    \label{eq:lbs}
\end{equation}
where $\mathbf{T} \in \mathbb{R}^{J \times 4 \times 4}$ is the joint transformations derived from $\hat{D}^t$, which are relative to the rest pose. By training the network modules to minimize the differences between the ground truth posed mesh $\mathbf{V}_d$ and the predicted deformed mesh $\hat{\mathbf{V}}_d$, the attention weights are optimized to function effectively as skinning weights.

\subsection{Training}
\label{sec:training}
\subsubsection{Dataset Preparation}
To prepare the dataset, we construct a motion database for each character model, following the procedures outlined in Lee et al.~\cite{lee2023same}. Specifically, the motion database is represented as follows:
\begin{equation}
    \mathcal{M} = \{ M_1, M_2, \dots, M_K \},
\end{equation}
where each motion clip $M_k = (S_k, D^{1:N_T}_k)$ consists of a skeleton $S_k$, which has a unique configuration, and corresponding motion data $D^{1:N_T}_k$. Starting with the initial skeletal motion data $M_1$, represented by skeleton $S_1$, which matches the size and body proportions of the character mesh, we augment the skeleton by randomly modifying its configuration. This involves adding or removing joints and adjusting bone lengths and root height of $S_1$.
This augmentation process results in a skeleton database:
\begin{equation}
    \mathcal{S} = \{ S_1, S_2, \dots, S_K \}.
\label{eq:skel-db}
\end{equation}
For each skeleton $S_k$ in this database, we retarget the motion data $D^{1:T}_1$ using the off-the-shelf retargeting method provided by MotionBuilder~\cite{autodesk2021motionbuilder}.
This results in a set of motion clips with diverse skeleton configurations, forming the complete motion database $\mathcal{M}$. Finally, we combine the mesh data and the motion database, yielding the character dataset $C = (G, \mathcal{M})$.

This process is repeated for all character models, producing a collection of character datasets:
\begin{equation}
    \mathcal{C}=\{C_1, C_2,..., C_{N_C}\},
\end{equation}
where ${N_C}$ is the number of character models. Because each character model has a distinct initial skeleton and mesh configurations, the number of vertices $N_V$ and the number of joints $N_J$ may differ between characters. However, $N_T$ remains consistent across the entire dataset because we use the same motion clips for all characters.

\subsubsection{Training Procedures}
During training, we randomly sample a pose at frame $t$ along with the corresponding geometry features, and we omit $t$ for brevity in this section. For the motion and mesh data, we randomly sample the input source pose $M_{src} = (S_{src}, D_{src})$ and the source mesh $G_{src} = \{ \mathbf{V}_r, \mathbf{V}_d, \mathbf{V}_f \}$. Notably, $M_{src}$ and $G_{src}$ do not need to originate from the same character model:
\begin{equation}
    M_{src} \in C_{i}, \quad G_{src} \in C_j, \quad i, j \in \{1, 2, \dots, {N_C}\},
\end{equation}
where $M_{src}$ is randomly sampled from the motion database $\mathcal{M}$, which is part of the character dataset $C_i$.
Given the sampled motion and mesh, our model predicts the deformed mesh vertices $\hat{\mathbf{V}}_d$ using Equation~\ref{eq:lbs}.

The objective terms to train the model are as follows:
\newcommand{\loss}[1]{\mathcal{L}_{#1}}
\begin{equation}
    \mathcal{L} = \loss{vtx} + \loss{edge} + \loss{skel} + \loss{sdf}.
\end{equation}
The vertex reconstruction loss $\loss{vtx}$ and edge loss $\loss{edge}$ measure the differences between the ground truth and predicted mesh in terms of deformed vertex positions and edges, respectively:
\begin{gather}
    \loss{vtx} = \frac{1}{N_V} \sum_{i=1}^{N_V} \lVert \mathbf{V}_{d, i} - \hat{\mathbf{V}}_{d, i} \rVert^2, \\
    \loss{edge} = \frac{1}{\lvert \mathcal{E} \rvert} \sum_{i, j \in \mathcal{E}} \lVert (\mathbf{V}_{d, i} - \mathbf{V}_{d, j}) - (\hat{\mathbf{V}}_{d, i} - \hat{\mathbf{V}}_{d, j}) \rVert^2,
\end{gather}
where $\mathcal{E}$ represents the set of edges connecting adjacent vertices and $\hat{\mathbf{V}}_d$ represents the deformed vertices predicted by our model. The edge loss ensures that the predicted mesh preserves the local rigidity of the ground truth mesh by accurately reconstructing local deformations along the edges. The skeleton loss $\loss{skel}$ and signed distance function~(SDF) loss $\loss{sdf}$ guide the learning of a target skeleton that aligns with the mesh, which are defined as follows:
\begin{gather}
    \loss{skel} = \mathrm{CD}(\mathbf{g}_{gt}, \mathbf{g}_{tgt}), \\
    \loss{sdf} = \mathrm{SDF}(\mathbf{V}_r, \mathbf{g}_{tgt}), 
\end{gather}
where $\mathbf{g}_{gt}$ and $\mathbf{g}_{tgt}$ represent the global offsets of the ground truth and predicted target skeleton, respectively.
$\mathrm{CD}(\mathbf{g}_{gt}, \mathbf{g}_{tgt})$ computes the Chamfer distance between these two, which is defined as follows:
\begin{equation}
    \mathrm{CD}(\mathbf{g}_{gt}, \mathbf{g}_{tgt}) =
    \frac{1}{\lvert \mathbf{g}_{gt} \rvert} \sum_{x \in \mathbf{g}_{gt}}{\min_{y \in \mathbf{g}_{tgt}}\lVert x-y \rVert^2_2}
    +
    \frac{1}{\lvert \mathbf{g}_{tgt} \rvert} \sum_{y \in \mathbf{g}_{tgt}}{\min_{x \in \mathbf{g}_{gt}}\lVert x-y \rVert^2_2}.
\end{equation}
The intuition behind using the Chamfer distance for the skeleton loss lies in its ability to compare the shapes of two different skeletons with varying numbers of joints by treating each joint as a point. Because the ground truth skeleton that fits the source mesh, which corresponds to $S_1$ of $C_j$, can possess a different number of joints from the target skeleton, we use the Chamfer distance to ensure that the predicted target skeleton has similar shape to the ground truth even when the number of joints and joint hierarchy are different. $\mathrm{SDF}(\mathbf{V}_r, \mathbf{g}_{tgt})$ computes the signed distance for each component of $\mathbf{g}_{tgt}$ with respect to the mesh at the rest pose $\mathbf{V}_r$. The SDF loss assigns negative values for joints inside the mesh and positive values for joints outside, encouraging the joints to be properly embedded within the character mesh.

One notable aspect of our approach is that we do not explicitly supervise the estimation of deformation parameters, such as skinning weights and bone offset residual factors. The reason for this is that the source skeletal structure $S_{src}$ can be arbitrary, with varying numbers of joints and hierarchical connections between them. As a result, it is infeasible to prepare a set of skinning weights that generalize across arbitrary skeletal structures. Instead, our method employs a self-supervised learning strategy through the vertex reconstruction loss. This encourages the model to implicitly learn the correct deformation parameters because the reconstruction loss can only be minimized when these deformation parameters are accurately estimated.

\section{Experiments}
\subsection{Implementation Details}
To obtain the character mesh data $G$, we randomly selected 17 characters for training and 9 characters for testing from Mixamo~\cite{mixamo}. For the motion data $M$, we used 14 motion sequences from the LaFAN1 dataset~\cite{harvey2020robust}, comprising a total of 75,351 frames sampled at 30fps. Using $G$ and $M$, we generated the deformed mesh vertices $\mathbf{V}_d$ via LBS, utilizing the skinning weights embedded in each Mixamo character. Additionally, to enhance training efficiency, we decimated the mesh vertices to fewer than 5,000 for characters with a higher vertex count.

Our method was implemented in PyTorch and executed on a single NVIDIA RTX A5000 GPU with 24GB of VRAM. We used the Adam optimizer~\cite{kingma2014adam}, with a fixed learning rate of $10^{-4}$ and a weight decay of $10^{-6}$. Each network module is followed by a batch normalization layer~\cite{ioffe2015batch} and the ReLU activation. The model was trained for 50 epochs with a batch size of 64 frames, which required approximately 30 minutes of computation per epoch.

\subsection{Baselines for Comparison}
We compared our method with other auto-rigging approaches, including Pinocchio~\cite{baran2007automatic}, RigNet~\cite{xu2020rignet}, and NBS ~\cite{li2021nbs}. We evaluated the performance of each method both qualitatively and quantitatively in terms of rigging, skinning, and mesh deformation. Pinocchio is the most relevant to our work because it generates plausible skeletal articulations and skinning weights based on the input skeleton and mesh. NBS also addresses a similar problem by generating skinning weights and mesh deformations given a mesh input, but it is limited by the fixed skeleton template. Consequently, we evaluated NBS only for skeletons that included all the joints corresponding to the predefined skeleton template. RigNet, a representative method for learning-based rig estimation, generates target skeletons for the given mesh without requiring skeleton inputs.

For evaluation, we utilized the augmented skeleton database of each character, which corresponds to $\mathcal{S}$ described in Equation \ref{eq:skel-db}, to validate the ability of our method to handle arbitrary skeletons and meshes. Each character was paired with its original skeleton $S_1$ and five additional augmented skeletons, denoted as $S_{2:6}$, which vary in number of joints and offset scales. To ensure a fair comparison across different methods, each of which has specific requirements for skeleton and mesh input, we took several steps to fully leverage this inhomogeneous skeleton dataset.

For Pinocchio, we pre-processed the character meshes to meet the requirement of being a single connected component, ensuring all vertices were linked within one graph. However, this process alters the number of vertices in the mesh, potentially affecting a fair comparison. To address this, we mapped the skinning weights predicted by Pinocchio back to the original mesh by finding the closest corresponding vertices. To deform the mesh with the predicted skeletal articulations and skinning weights, we utilized the SAME model to retarget the source motion to the predicted skeleton. RigNet does not allow for explicitly specifying the desired number of joints, but it can generate multiple skeletons from a single input mesh. Therefore, we randomly generated five different skeletons using RigNet for evaluation, addressing variability in shape and structure of the generated skeletons. To ensure a fair comparison, we evaluated NBS only on $S_1$, which is the skeleton configuration that meets the requirements of NBS across all characters, while we evaluated Pinocchio and our method on all skeleton variations, $S_{1:6}$. Throughout the evaluations of RigNet and NBS, we used the pre-trained model provided by the authors.

\subsection{Evaluation}
\begin{figure*}[t]
    \centering    \includegraphics[width=\textwidth]{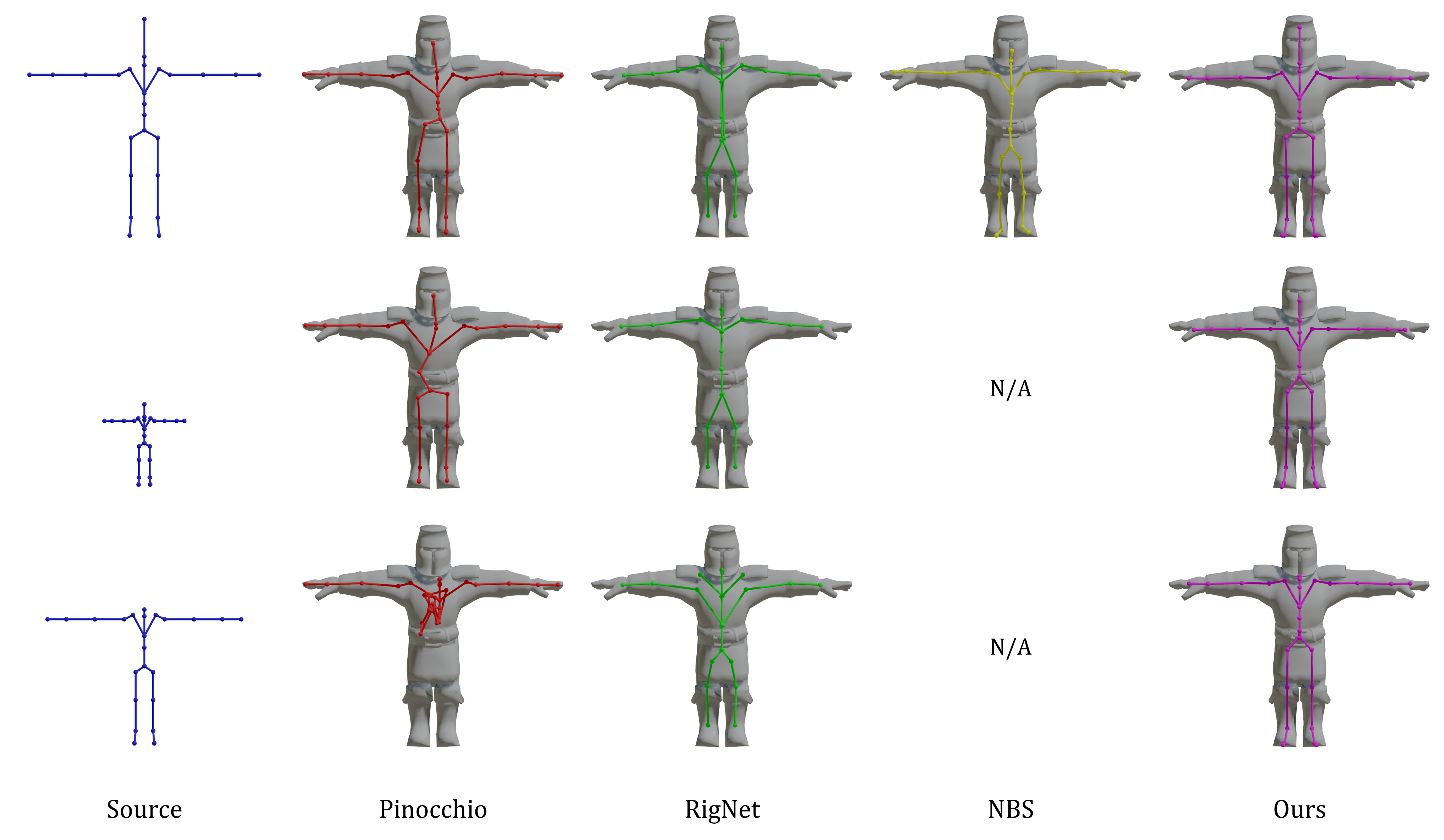}
    \caption{Comparison to baselines on skeleton prediction results given the same mesh with different source skeletons. Each skeleton has distinct body scales and bone lengths, with varying numbers of joints: from the top 25, 24, and 23 joints.}
    \label{fig:qual_rigging}
    \vspace{-1em}
\end{figure*}
\noindent{\textbf{Rigging}} \hspace{0.5mm}
To evaluate rigging quality, we compared the similarity between the predicted and ground truth skeletons, which corresponds to $S_1$ for each character. The ground truth skeletons were derived from the character models obtained from Mixamo~\cite{mixamo}. For evaluation, we employed three Chamfer distance-based metrics, making them suitable for comparing skeletons with different numbers of joints: CD-J2J (joint-to-joint), CD-J2B (joint-to-bone), and CD-B2B (bone-to-bone)~\cite{xu2020rignet}. These metrics quantify the spatial discrepancies between the predicted and ground truth skeletons, where lower values indicate a closer alignment between the predicted and ground truth skeletal shapes. For more details, please refer to RigNet~\cite{xu2020rignet}. 

Table \ref{table:rigging} presents that our method yielded better results than the baselines across most metrics. While RigNet achieved slightly better results in CD-B2B than ours, the difference was marginal compared to the differences observed in other metrics. Moreover, considering that our approach enables greater flexibility in handling diverse skeleton and mesh configurations while maintaining the integrity of the given skeletal structure, these superior results are highly significant.

Figure~\ref{fig:qual_rigging} shows that our method outperformed the baselines with significantly higher flexibility in the number of joints and bone lengths. Pinocchio struggled to preserve structural integrity given discrepancies in size and proportion between the skeleton and mesh, leading to asymmetric joint predictions along the up-axis. RigNet faced difficulties in accurately identifying the pelvis joint, and introduced unnecessary joints in bulky regions, such as the shoulders, causing the skeleton to exhibit significant deviations from the source skeletal structure. While NBS successfully generated skeletons that fit the size and proportions of the mesh, it altered the source skeleton shape into predefined skeletal offsets, which resulted in deviations from the source. Furthermore, NBS was not operable when the source skeleton did not contain the joints defined in the predefined template, as indicated by the N/A in the last two rows of Figure \ref{fig:qual_rigging}. In contrast, our method successfully predicted the target skeleton, showing consistent rig predictions even in cases with substantial variations in size and proportion between the mesh and skeleton.

\begin{table} [t!]
    \centering
    \caption{Quantitative results on rigging prediction. The best result for each column is in bold.} 
    \begin{tabular}{cccc}
    \hline
              & CD-J2J↓          & CD-J2B↓          & CD-B2B↓       \\ \hline
    Pinocchio & 34.03            & 26.81            & 25.10         \\ 
    RigNet    & 18.88            & 13.01            & \textbf{8.64} \\ 
    NBS       & 20.15            & 15.70            & 11.92         \\ \hline
    Ours      & \textbf{15.25}   & \textbf{10.87}   & 9.07          \\ \hline
    \end{tabular}
    \label{table:rigging}
    \vspace{-1em}
\end{table}

\begin{figure*}[t!]
    \centering    \includegraphics[width=\textwidth]{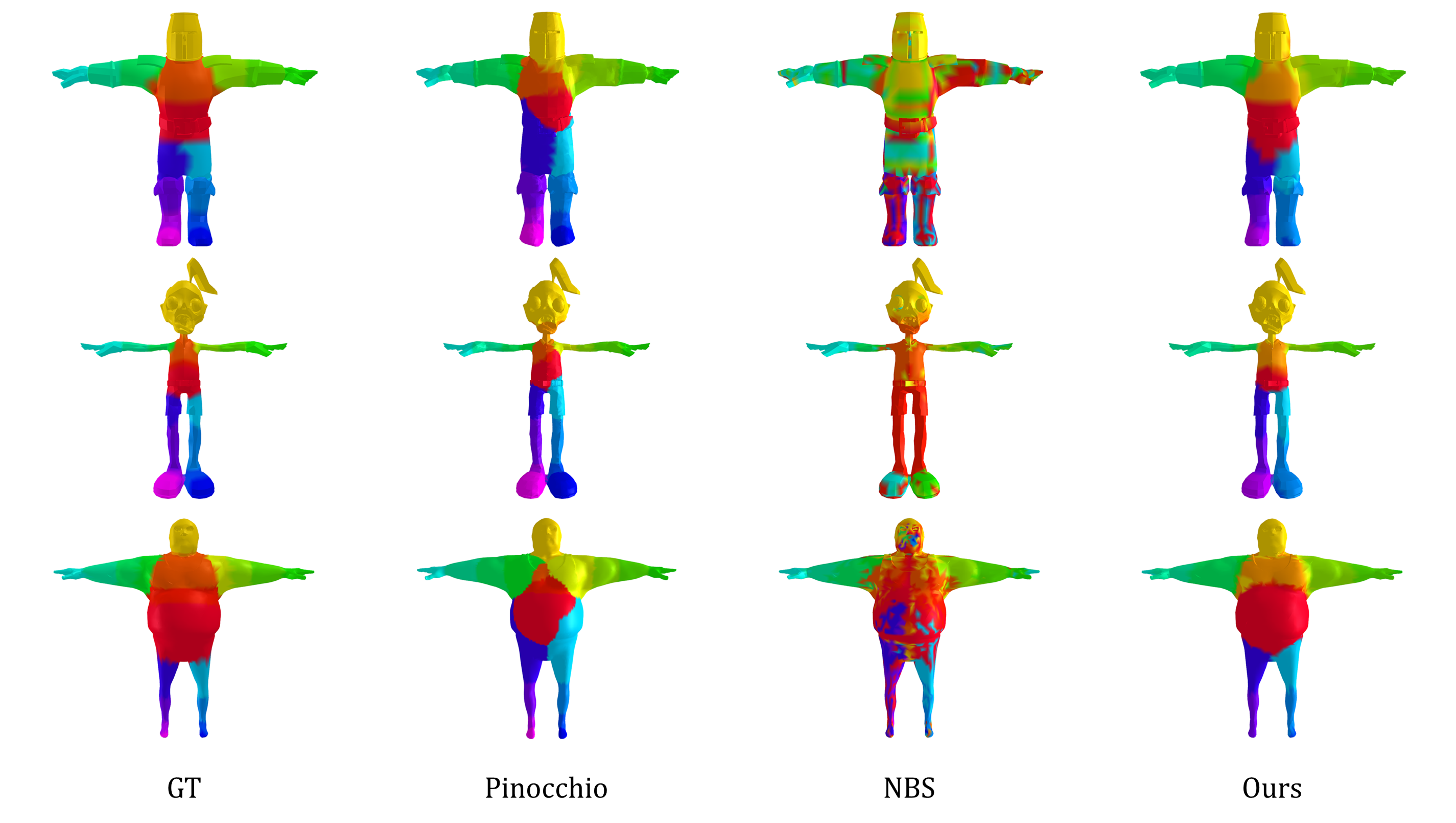}
    \caption{Skinning weight results predicted from a source skeleton with a fixed number of joints.}
    \label{fig:qual_skinning_fixed}
    \vspace{-1em}
\end{figure*}
\noindent{\textbf{Skinning and Deformation}} \hspace{0.5mm}
To evaluate the accuracy of skinning weight prediction, we measured the difference between predicted and ground truth skinning weights by computing the average L1 distance (Skinning L1) between them. Because the ground truth skinning weight is only available for $S_1$, we measured the L1 distance exclusively for $S_1$, while other metrics were measured across all skeleton variations $S_{1:6}$. Additionally, to evaluate the deformation quality driven by the predicted skinning weights, we measured the Chamfer distance (CD), average distance error (ADE), and max distance error (MDE) between the predicted and ground truth deformed meshes. These metrics were measured between vertices deformed using the predicted skinning weights and those deformed with the ground truth weights. To further evaluate the smoothness of the deformed meshes, we measured Edge Length Score (ELS)~\cite{wang2023zpt}. This score compares the length of each edge in the predicted mesh to that of the corresponding edge in the ground truth mesh, which is calculated as follows:
\begin{gather}
    \mathrm{ELS}(\mathbf{V}_d, \mathbf{\hat{V}}_{d}) = \frac{1}{\lvert \mathcal{E} \rvert}\sum_{\{i,j\}\sim \mathcal{E}}^{}1-\left| \frac{\left\| \mathbf{\hat{V}}_{d, i}-\mathbf{\hat{V}}_{d, j} \right\|_2}{\left\| \mathbf{V}_{d, i}-\mathbf{V}_{d, j} \right\|_2} -1\right|, 
\end{gather}
where $\mathcal{E}$ denotes the entire edges of the mesh, $\hat{\mathbf{V}}_{d, i}$ and $\hat{\mathbf{V}}_{d, j}$ are the predicted vertices in the deformed mesh, and $\mathbf{V}_{d, i}$ and $\mathbf{V}_{d, j}$ are the vertices in the ground truth mesh. 

\begin{table}
    \centering
    \caption{Quantitative results on skinning and deformation. The best result for each column is in bold.}
    \begin{tabular}{cccccc}
    \hline
              & \makecell{Skinning \\ L1↓} & CD↓           & ADE↓          & MDE↓           & ELS↑          \\ \hline
    Pinocchio & \textbf{0.0236}            & 15.14         & 16.78         & 52.68          & \textbf{0.94} \\ 
    NBS       & 0.0599                     & 18.11         & 31.60         & 154.40         & -5.65         \\ \hline
    Ours      & 0.0449                     & \textbf{7.15} & \textbf{8.09} & \textbf{26.18} & 0.89          \\ \hline
    \end{tabular}
    \label{table:quan_skinning}
    \vspace{-1em}
\end{table}
\begin{figure}[t]
    \centering
    \includegraphics[width=\columnwidth]{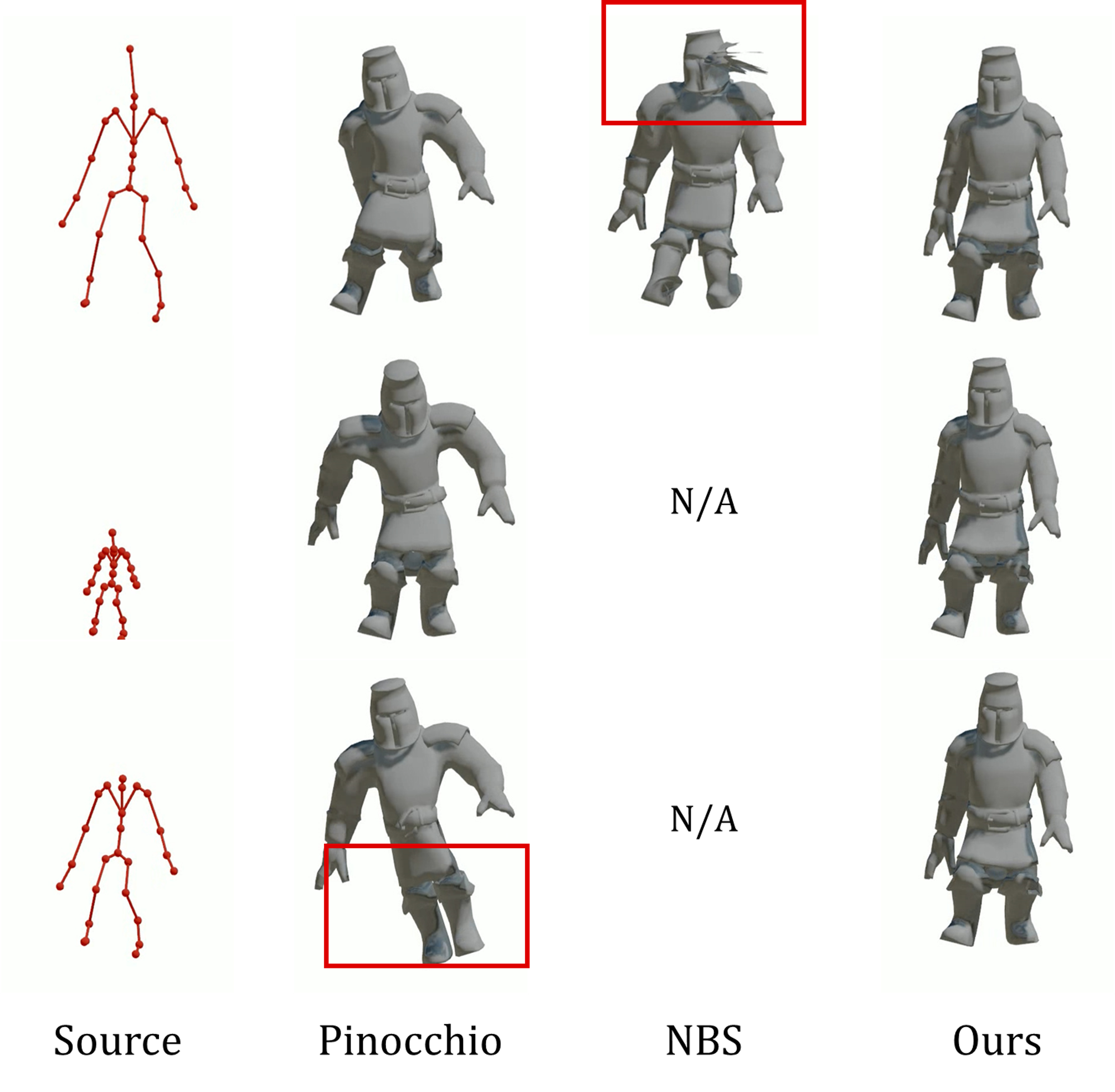}
    \caption{Qualitative comparison with baselines on mesh deformation.}
    \label{fig:qual_comparison_deformed}
    \vspace{-1em}
\end{figure}
\begin{figure*}[t]
    \centering    \includegraphics[width=\textwidth]{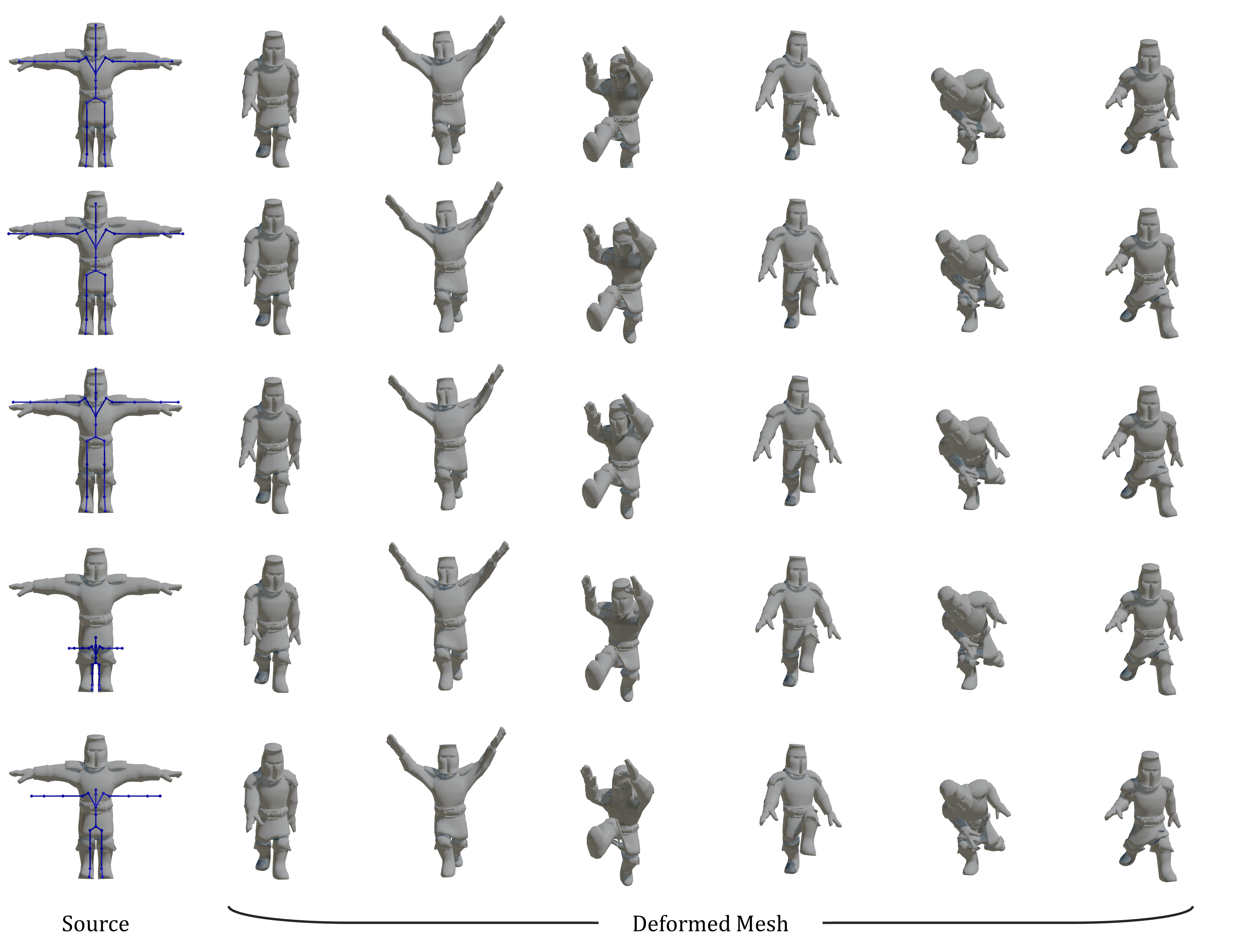}
    \caption{Given the same source mesh with different source skeletons, our method robustly generates consistently deformed meshes in accordance with skeletal movements, even when the source skeletons have diverse configurations with varying number of joints: from the top 25, 22, 20, 24, and 23 joints are respectively used. The target skeletons derived from the source are presented in the last column of Figure~\ref{fig:qual_rigging}.}
    \label{fig:qual_deformed}
    \vspace{-1em}
\end{figure*}
\begin{table}
    \centering
    \caption{Deformation metrics with different skeletal configurations. $S_1$ represents the case in which the input skeleton is exactly aligned with the mesh, while $S_{2:6}$ represents cases with augmented skeletons.
    Despite the misalignment, $S_{2:6}$ still produces competitive results, demonstrating the robustness of our method.}
    \begin{tabular}{ccccc}
    \hline
              & CD↓  & ADE↓ & MDE↓   & ELS↑ \\ \hline
    $S_1$     & 7.15 & 8.09 & 26.18  & 0.89 \\ 
    $S_{2:6}$ & 7.54 & 8.63 & 28.38  & 0.89 \\ \hline
    \end{tabular}
    \label{tab:deform-different-skeletons}
    \vspace{-1em}
\end{table}
Table \ref{table:quan_skinning} shows the quantitative results on skinning and deformation.
In Skinning L1, Pinocchio outperformed ours. The suboptimal performance of ours can be attributed to the challenge of predicting accurate skinning weights for joints with minimal movements, such as the chest, especially without ground truth supervision during training. Specifically, these skinning weights are indirectly optimized through self-supervised learning, which can be less accurate when there are no significant vertex displacements associated with those joints. Nonetheless, our method consistently produced visually plausible skinning weights that aligned well with the relevant body parts, as shown in Figure \ref{fig:qual_skinning_fixed}. While Pinocchio produced reasonable results for the arms and head, it incorrectly assigned irrelevant vertices from the lower body and pelvis joint to other joints. This is because Pinocchio relies on geometric processing for skeleton estimation, which does not account for the high-level semantic relationships that define which vertices should be bound to specific joints. On the other hand, NBS produced unstable skinning predictions with inconsistent results across body parts. This is because NBS requires the source mesh to be strictly aligned with the SMPL distribution due to its dependency on the training dataset, limiting its performance on stylized, non-human characters. In contrast, ours demonstrated consistent and accurate skinning weight predictions that closely resemble the ground truth, indicating its ability to effectively capture the relationships between mesh vertices and skeleton joints.

When evaluating deformation metrics across all skeleton variations $S_{1:6}$, our method outperformed the baselines in CD, ADE, and MDE, as shown in Table~\ref{table:quan_skinning}, although the ELS results were slightly lower than those of Pinocchio. Additionally, Figure~\ref{fig:qual_comparison_deformed} presents the visual results of deformed meshes based on the predicted skeletons shown in Figure~\ref{fig:qual_rigging}. While Pinocchio produced plausible results in general, it failed to retarget the given pose when the predicted skeleton significantly deviated from the shape of the source skeleton. Similarly, although NBS predicted a plausible target skeleton, unstable skinning weights employed by NBS affected the quality of the mesh deformation. In contrast, ours robustly predicted the target skeleton given variations of source skeletons and generated consistently deformed results across varying skeletal configurations. For animation results, please see the supplementary video.
\begin{figure}[t]
    \centering    \includegraphics[width=\columnwidth]{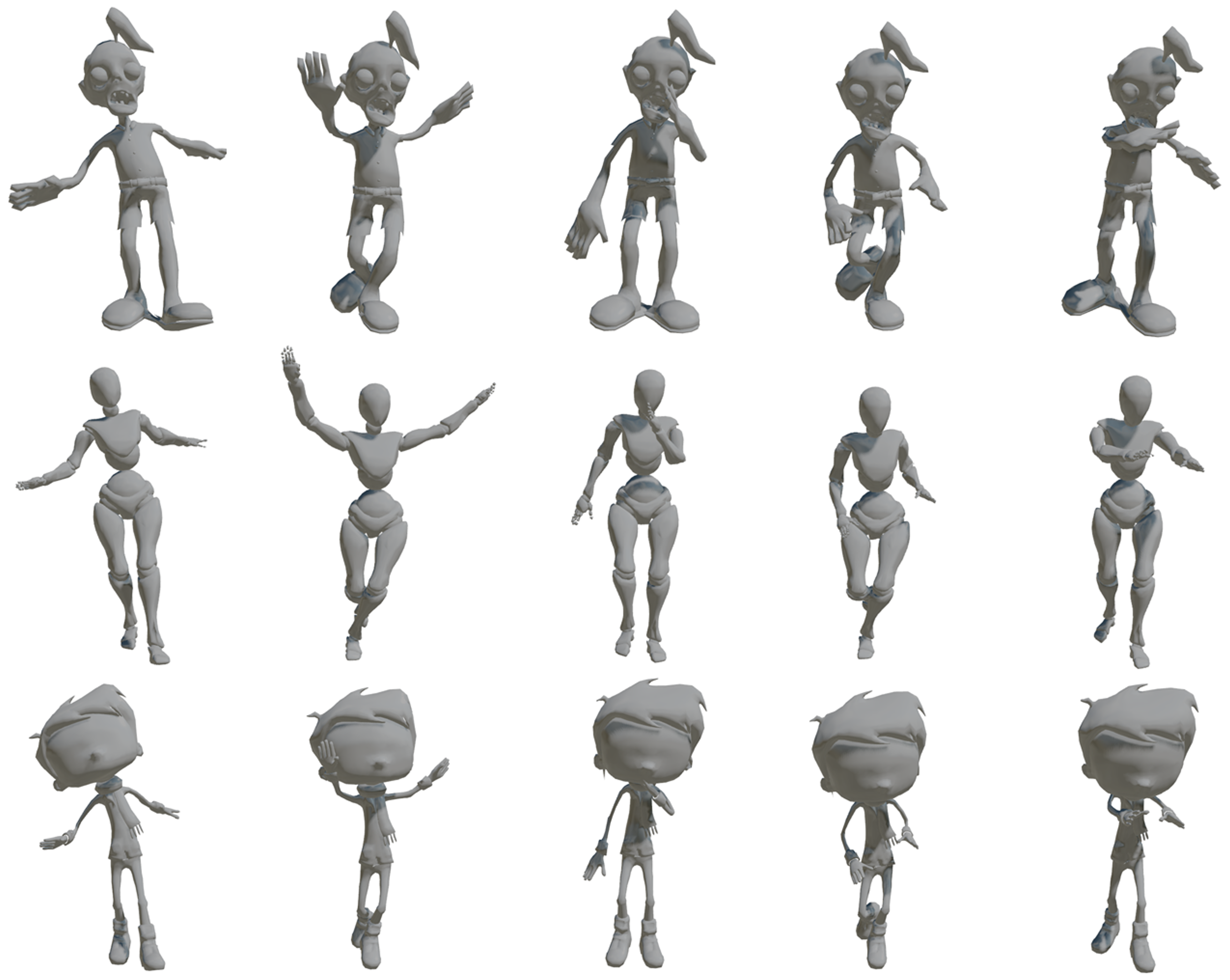}
    \caption{Deformation results for multiple characters in the same pose. Each row shows different characters, while each column represents the same input pose applied to all characters.}
    \label{fig:qual_deformed_multichar} 
    \vspace{-1.5em}
\end{figure}
To evaluate the robustness of our method in handling diverse skeletal configurations for a given input mesh, we measured the deformation metrics in two scenarios: (i) when the input skeleton is perfectly aligned with the mesh, which corresponds to $S_1$, and (ii) when the input skeleton differs from the mesh, which corresponds to $S_{2:6}$. As presented in Table \ref{tab:deform-different-skeletons}, $S_1$ consistently produced better deformation compared to $S_{2:6}$, indicating that a more accurate initial alignment of the skeleton yields improved results. However, the performance gap between $S_1$ and $S_{2:6}$ was not substantial. Furthermore, even the mismatched skeletons $S_{2:6}$ produced outperforming or comparable results to baselines whose results are reported in Table \ref{table:quan_skinning}. This result highlights the robustness of our method on adaptive rigging and skinning even when the input skeleton does not perfectly align with the mesh. The visual results are presented in Figure~\ref{fig:qual_deformed}, where the same source mesh, paired with different skeleton configurations, produced consistently deformed results. In addition, Figure~\ref{fig:qual_deformed_multichar} demonstrates that our method reliably produces consistent poses across different character meshes. For animation results, please see the supplementary video. 
\subsection{Ablation Studies}
\begin{figure} [t!]
    \centering    \includegraphics[width=\columnwidth]{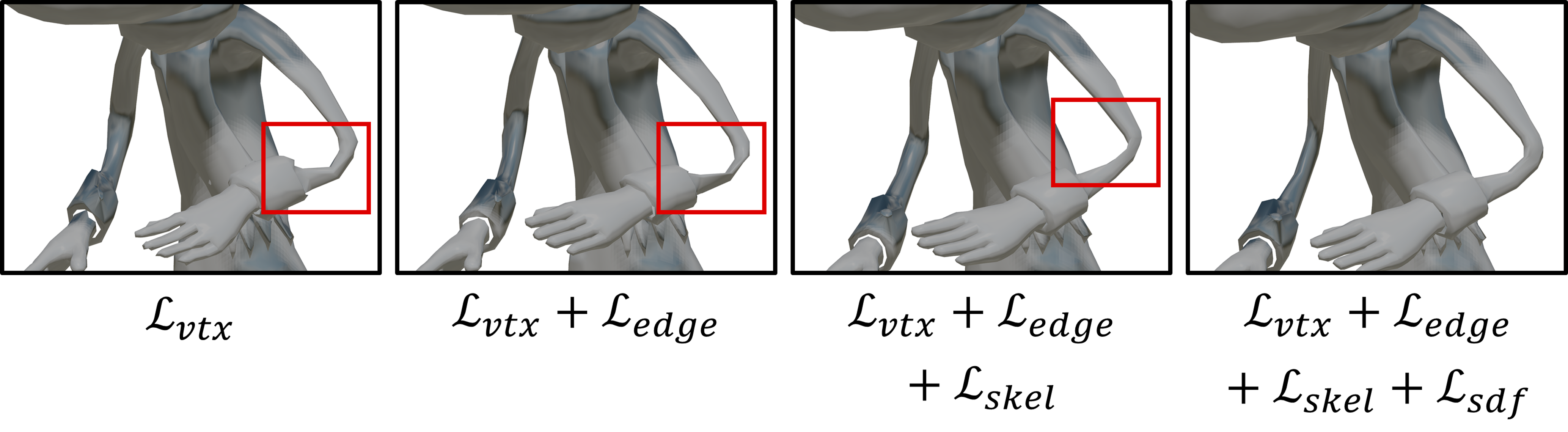}
    \caption{Qualitative results of the ablation study for each loss term.}
    \label{fig:ablation_loss} 
    \vspace{-1em}
\end{figure}

To evaluate the contribution of each loss term, we conducted an ablation study by progressively removing individual loss terms, and the results are shown in Table \ref{tab:ablation} and Figure \ref{fig:ablation_loss}. We observed that the edge loss $\mathcal{L}_{edge}$ consistently improved the evaluation results of all the metrics compared to the case of training solely with $\mathcal{L}_{vtx}$, showing its contribution to enhanced overall performance. Notably, the models trained with $\mathcal{L}_{vtx}$ and $\mathcal{L}_{vtx}+\mathcal{L}_{edge}$ yielded slightly better quantitative results in terms of deformation, measured by CD, ADE, and MDE, compared to the full model. However, their quantitative results on rigging reflected in CD-J2J, CD-J2B, and CD-B2B, were significantly worse. This suggests that optimizing solely on deformation-related objectives does not adequately account for skeletal articulations, leading to distortions like candy-wrapper artifacts around the arm joints, as shown in Figure \ref{fig:ablation_loss}. Therefore, we emphasize the significance of rigging-related loss terms, which are $\mathcal{L}_{skel}$ and $\mathcal{L}_{sdf}$, for improving the overall deformation quality.

The skeleton loss $\mathcal{L}_{skel}$ significantly enhanced the rigging quality, as indicated by all values of CD-J2J, CD-J2B, and CD-B2B, demonstrating its effectiveness in aligning character rigs with the mesh proportions. Additionally, it mitigated deformation artifacts, leading to smoother mesh shapes compared to the models trained with deformation objectives alone. The SDF loss $\mathcal{L}_{sdf}$ enhanced both rigging accuracy and deformation quality, with the full loss combination yielding the best results. These findings indicate the importance of using all the loss terms to achieve the best rigging results without compromising the deformation quality. 

\begin{figure} [t!]
    \centering    \includegraphics[width=\columnwidth]{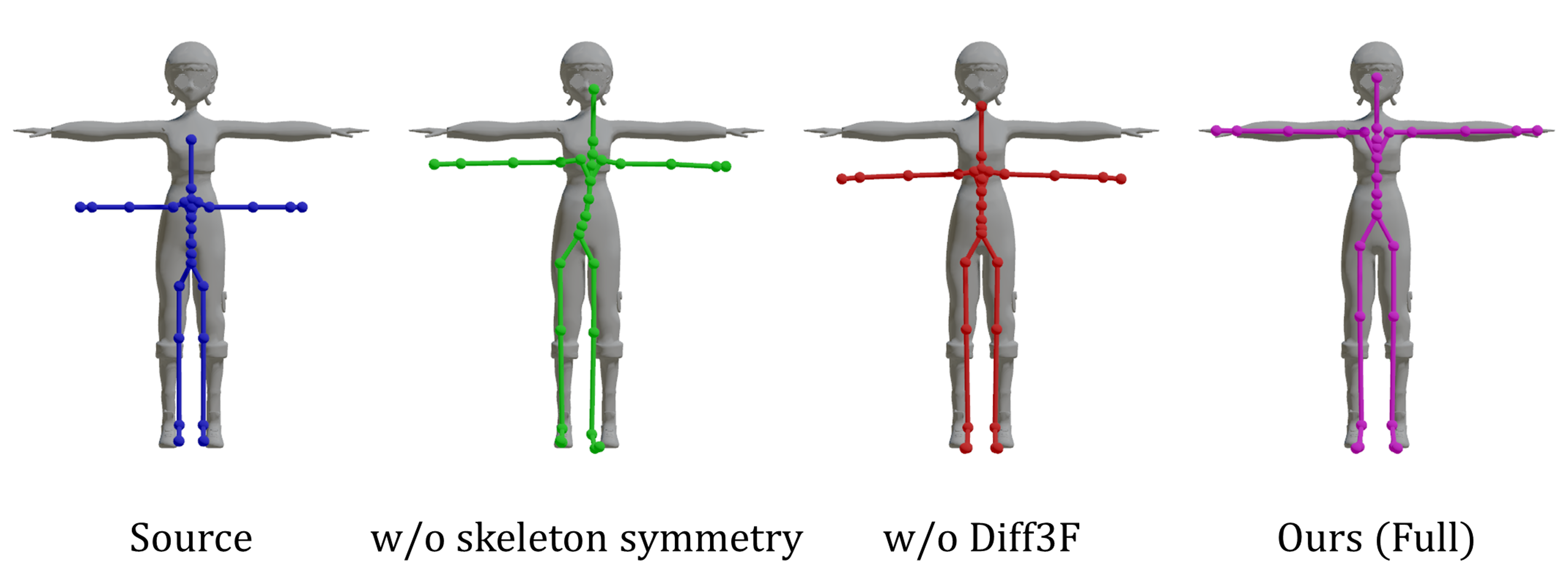}
    \caption{Ablation on target skeleton prediction.}
    \label{fig:ablation_rigging}
\end{figure}
\begin{figure} [t!]
    \centering    \includegraphics[width=\columnwidth]{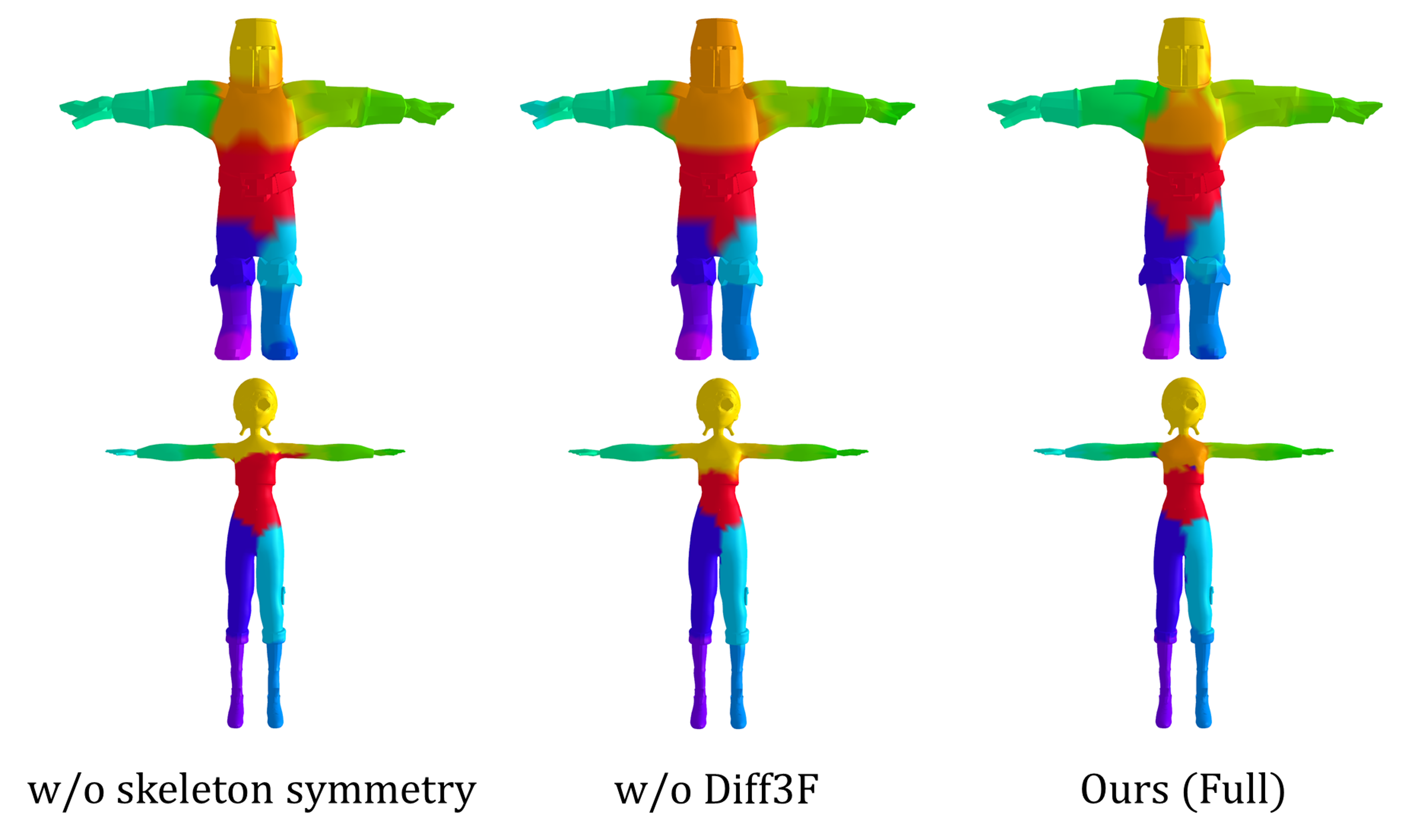}
    \caption{Ablation on skinning prediction.}
    \label{fig:ablation_skinning}
    \vspace{-1em}
\end{figure}
\begin{table*}[t!]
\centering
\caption{Ablation study results with a source skeleton having \textbf{varying} number of joints and \textbf{differing} size and proportion from the source mesh.}
\begin{tabular}{cccc|ccccc}
\hline
\multicolumn{1}{l}{Metric}                                        & CD-J2J↓        & CD-J2B↓        & CD-B2B↓       & Skinning L1↓    & CD↓           & ADE↓    & MDE↓     & ELS↑          \\ \hline
\multicolumn{1}{l}{\( \mathcal{L}_{vtx} \)}                       & 18.82          & 14.26          & 11.59         & 0.0461          & 7.54          & 8.63          & 27.98          & 0.87          \\
\multicolumn{1}{l}{\( \mathcal{L}_{vtx} + \mathcal{L}_{edge} \)}  & 17.99	       & 13.52          & 10.70         & 0.0449          & \textbf{7.39} & \textbf{8.41} & \textbf{27.63} & 0.88          \\
\multicolumn{1}{l}{\( \mathcal{L}_{vtx} + \mathcal{L}_{edge}
                    + \mathcal{L}_{skel}\)}                       & 16.54          & 12.52          & 10.37         & \textbf{0.0418} & 7.80          & 9.04          & 29.86          & 0.88          \\ 
\multicolumn{1}{l}{\( \mathcal{L}_{vtx} + \mathcal{L}_{edge}
                    + \mathcal{L}_{skel} + \mathcal{L}_{sdf}\)} & \textbf{16.28} & \textbf{11.85} & \textbf{9.90} & 0.0449          & 7.48          & 8.54          & 28.01          & 0.89          \\ \hline
\multicolumn{1}{l}{w/o Diff3F}                                    & 16.43          & 12.16          & 10.23         & 0.0503          & 7.56          & 8.69          & 29.60          & \textbf{0.91} \\ 
\multicolumn{1}{l}{w/o skeleton symmetry}                         & 17.90          & 13.70          & 11.41         & 0.0505          & 8.14          & 9.52          & 31.86          & 0.90          \\ \hline
\end{tabular}
\label{tab:ablation}
\vspace{-1em}
\end{table*}
We also analyzed the impact of Diff3F and skeleton symmetry by excluding each component.
As shown in Figure~\ref{fig:ablation_rigging}, while our full model produced accurate rigging results, with joints precisely embedded within the character mesh, the model trained without Diff3F produced joints that deviate from the character mesh, emphasizing its role in maintaining joint-vertex correspondences. Moreover, the model trained without skeletal symmetry processing struggled to position the joints correctly in relation to the mesh. We also present the importance of both components for accurate character skinning in Figure~\ref{fig:ablation_skinning}. Excluding Diff3F led to the leakage of skinning weights to unintended body parts, demonstrating the importance of Diff3F in capturing accurate joint-vertex correspondences. Additionally, removing skeletal symmetry produced asymmetric skinning weights, which could cause inaccurate deformations when the character poses are applied. These results are also numerically evident in Table \ref{tab:ablation}, with the worst scores for the Skinning L1. Overall, these results validate the effectiveness of incorporating both Diff3F and skeletal symmetry to improve the quality of rigging and skinning, which is crucial for high-quality mesh deformations. For animation results, please see the supplementary video.
\begin{figure} [t!]
    \centering    \includegraphics[width=\linewidth]{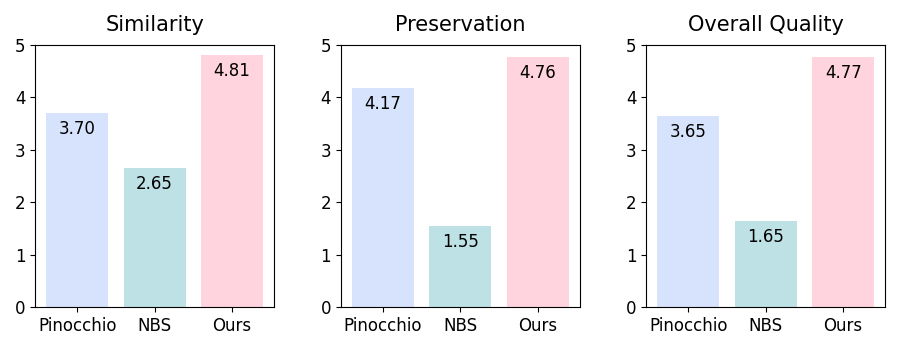}
    \caption{Results of user study in terms of Similarity, Preservation, and Overall Quality.}
    \label{fig:userstudy}
    \vspace{-2em}
\end{figure}
\subsection{User Study}
We conducted a user study to evaluate the deformation quality resulted from the predicted skeleton and skinning weights of our method. We compared our approach with other skeleton-based auto-rigging methods: Pinocchio and NBS. The study was designed to compare the quality of the deformed mesh animations generated by transferring a given source skeletal animation to a source mesh in the rest pose. We sampled 11 source skeleton-mesh pairs from our test dataset, including augmented skeletons. As a result, we generated a total of 33 tasks, each comprising 4-5 seconds of animations created by each method. For each resulting mesh animation that is presented in a random order, participants were asked to evaluate the following three criteria: \textit{Similarity} with the source skeletal animation, \textit{Preservation} of the original mesh shape compared to its rest pose state, and \textit{Overall Quality} of the animation. All questions were rated on a 5-point Likert scale, ranging from 1 (Strongly disagree) to 5 (Strongly agree). We recruited 16 participants (9 males and 7 females; ages: 24 to 33).

The results of the study are shown in Figure \ref{fig:userstudy}.
Our method achieved the highest scores for all three criteria, with 4.81 in \textit{Similarity}, 4.76 in \textit{Preservation}, and 4.77 in \textit{Overall Quality}, demonstrating its superior performance in both rigging and skinning quality. Pinocchio scored high in \textit{Preservation} but showed relatively lower performance in the other two criteria. This low performance is due to the suboptimal rigging quality created for the augmented skeletons, as shown in Figure \ref{fig:qual_rigging}. Specifically, the predicted skeleton frequently collapsed into a localized region of the mesh, resulting in limited deformation across the remaining body parts. NBS produced even lower scores, particularly in \textit{Preservation} and \textit{Overall Quality}, due to the presence of noticeable artifacts such as abnormal stretching, as shown in Figure \ref{fig:qual_comparison_deformed}. This along with the inaccurate prediction of skinning weights shown in Figure \ref{fig:qual_skinning_fixed} also negatively impacted the \textit{Similarity} score, making it relatively lower compared to other methods. In contrast, the robust performance of our method in predicting both skeleton and skinning weights, even with variations in skeleton configurations, resulted in high-quality pose-conditioned mesh deformations that reflect the naturalness on human perception.

\section{Discussion and Conclusion} 
\begin{figure}
    \centering    \includegraphics[width=\columnwidth]{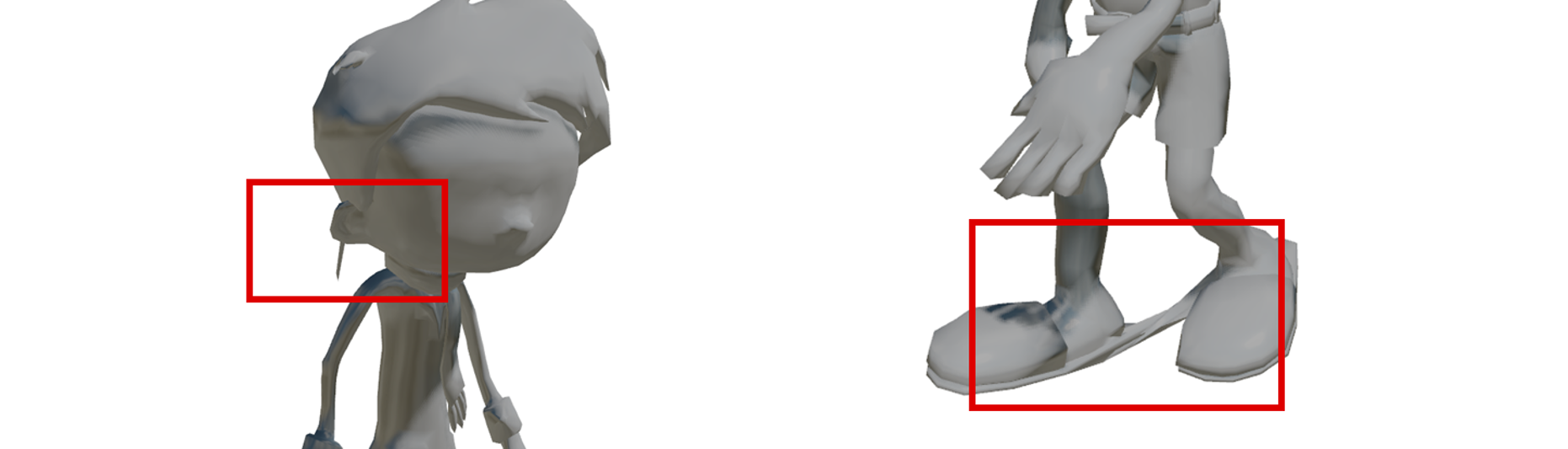}
    \caption{When the skinning weights are inaccurately predicted, artifacts may appear during the deformation process.}
    \label{fig:limitation}
    \vspace{-1em}
\end{figure}
\begin{figure*} 
    \centering
    \begin{minipage}{0.45\textwidth}
        \centering
        \includegraphics[width=\textwidth]{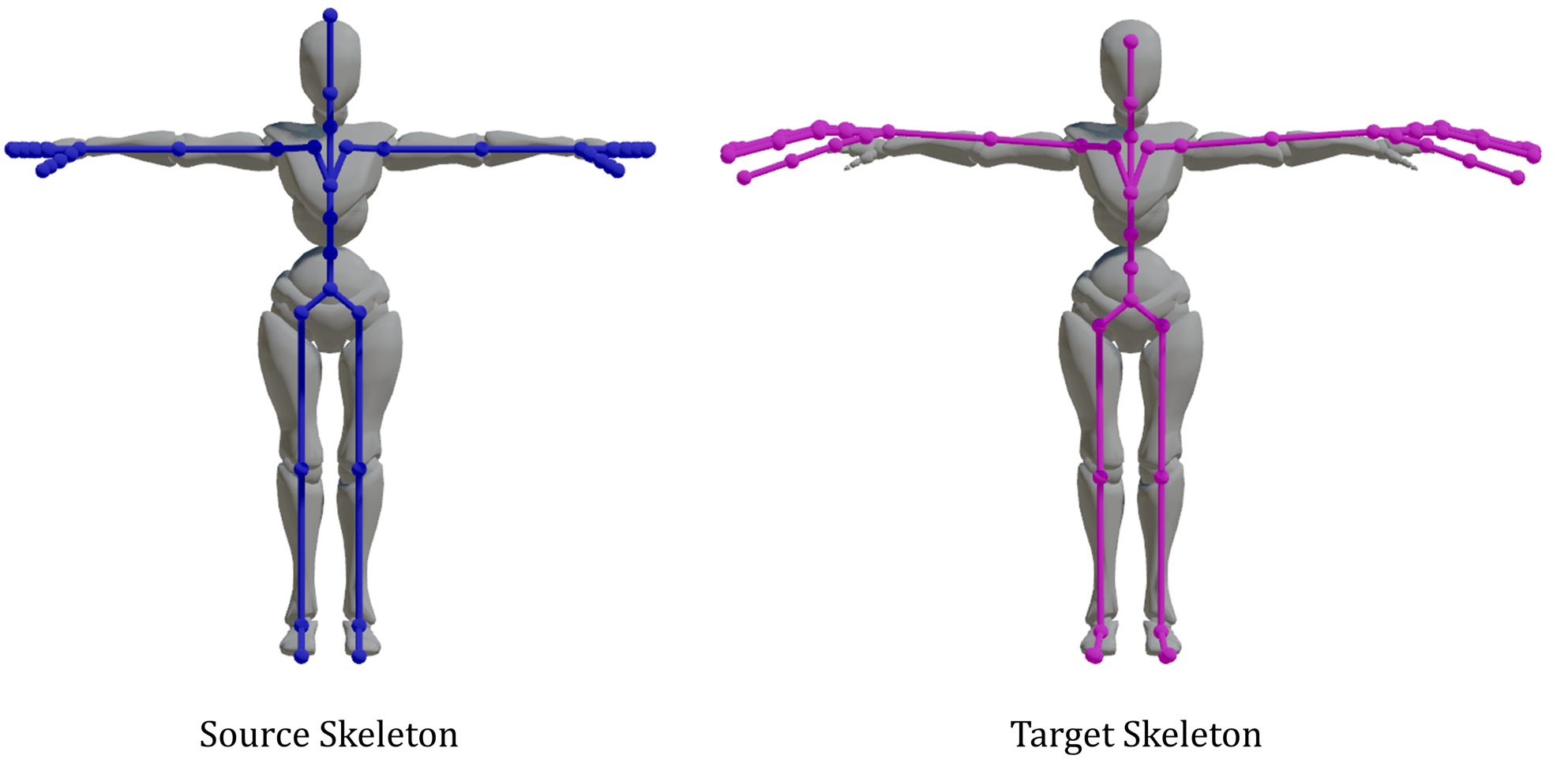}
        \subcaption{}        
        \label{fig:limitation_skel_with_fingers_a}
    \end{minipage}
    \begin{minipage}{0.31\textwidth}
        \centering
        \includegraphics[width=\textwidth]{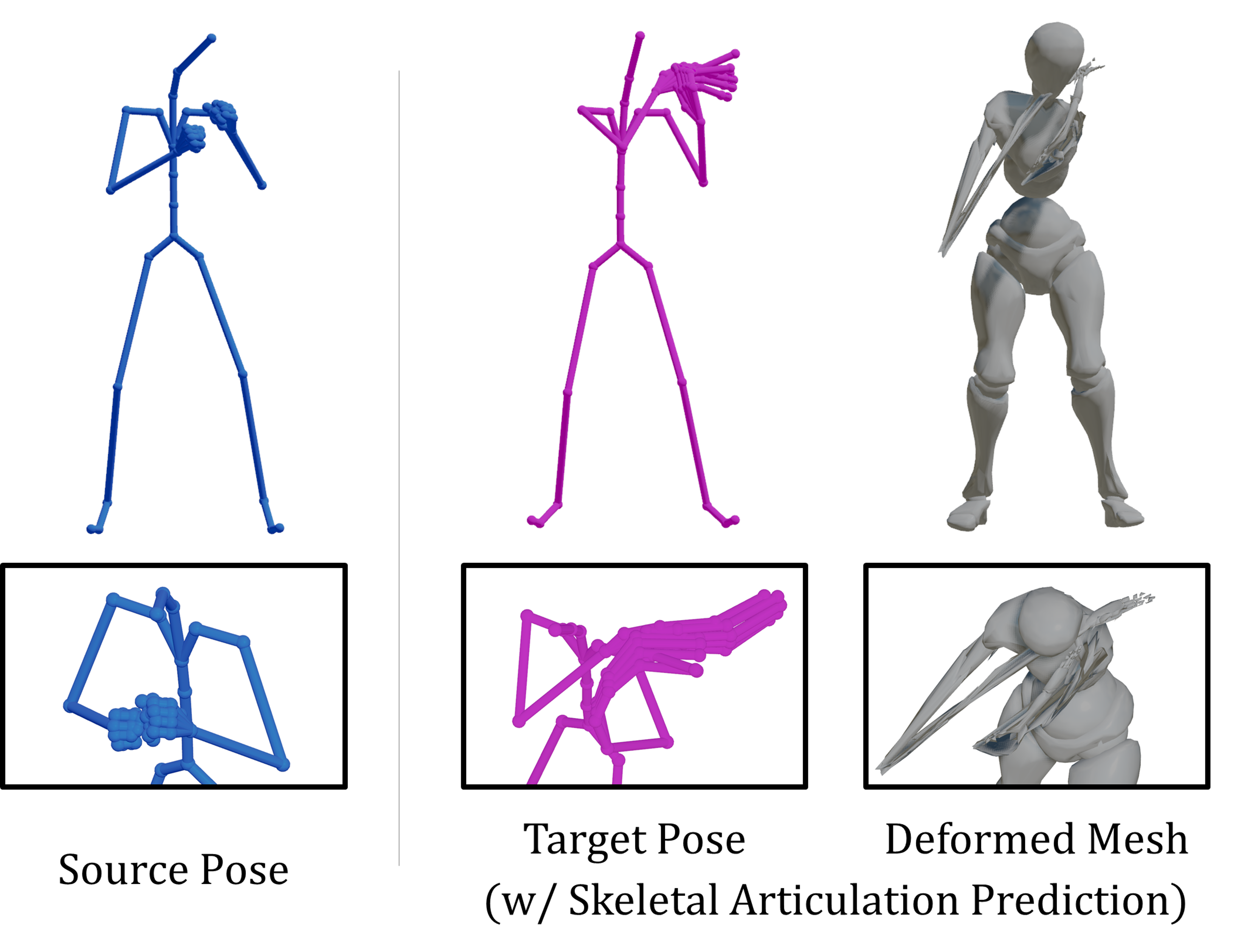}
        \subcaption{}
        \label{fig:limitation_skel_with_fingers_b}
    \end{minipage}
    \begin{minipage}{0.20\textwidth}
        \centering 
        \includegraphics[width=\textwidth]{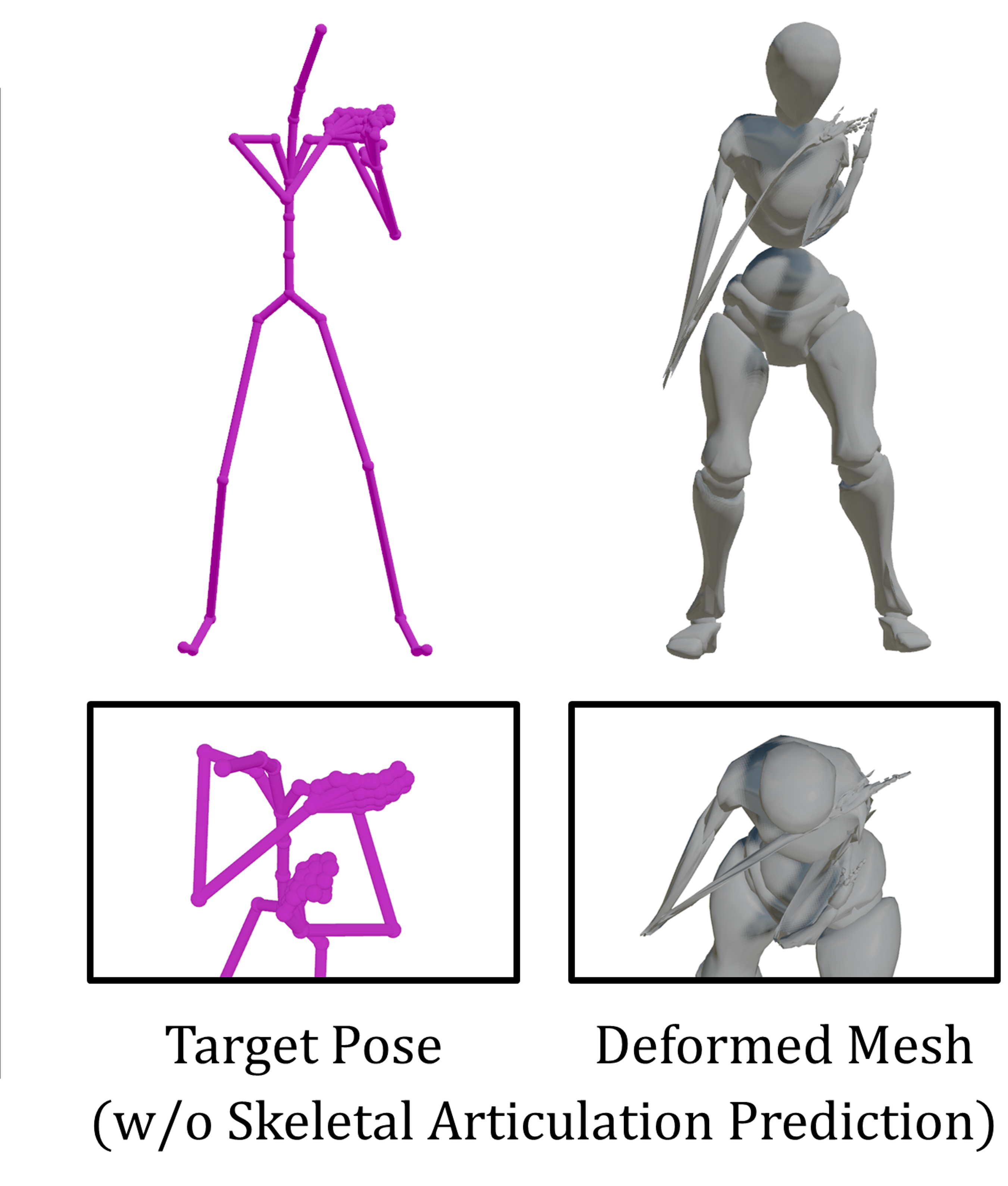}
        \subcaption{}
        \label{fig:limitation_skel_with_fingers_c}
    \end{minipage}
    \caption{(a) Source skeleton with finger joints (left) and the target skeleton predicted by the Skeletal Articulation Prediction module (right). Skeletons are overlapped on source meshes to demonstrate alignments between them. (b) Target pose and deformed mesh obtained by retargeting the source pose to the predicted target skeleton. (c) Target pose and deformed mesh obtained by directly using the source skeleton as the target skeleton without Skeletal Articulation Prediction.}
    \label{fig:limitation_skel_with_fingers}
    \vspace{-1em}
\end{figure*}
Despite the flexibility and high-quality results, we observed some limitations of our method. While our method consistently produced plausible results across diverse mesh and skeleton variations, there was occasional leakage of skinning weights, due to the bounding of vertices to unintended joints. Furthermore, the predicted skinning weights were sometimes distributed evenly across multiple joints, instead of being concentrated on a few specific joints, which can cause indistinct deformations that fail to follow the given pose adequately. This can result in artifacts in the deformed meshes, as shown in Figure~\ref{fig:limitation}, significantly reducing the overall deformation quality. Additionally, our method struggles to handle auxiliary joints, such as fingers. As shown in Figure~\ref{fig:limitation_skel_with_fingers_a}, the predicted target skeleton deviates significantly from the source mesh, resulting in undesirable mesh deformation caused by inaccurate motion retargeting and skinning prediction, as illustrated in Figure~\ref{fig:limitation_skel_with_fingers_b}. Even when the source skeleton is directly given to the Skinning Weight Prediction module, the deformation still yields unnatural results, as shown in Figure~\ref{fig:limitation_skel_with_fingers_c}. This issue stems partly from the limitations of SAME in accurately retargeting motions for auxiliary joints, and partly from the difficulty in predicting appropriate skinning weights for finer skeletal joints like fingers. Finally, to compute the SDF loss, we first obtain unsigned distances from the joints to the nearest surface, and then assign signs based on inside and outside classification, enabling SDF computation for non-watertight meshes. However, this simplified approach lacks robustness, particularly for complex geometries.

For future work, we aim to improve rigging quality by ensuring that each joint is exactly embedded within the character mesh. Enhancing the accuracy of motion retargeting by incorporating more advanced techniques and utilizing more robust prior features will also be an interesting direction to improve performance. Additionally, incorporating additional regularization to encourage sharp changes in skinning weights between different joints may further enhance deformation quality.

In this paper, we introduced a novel end-to-end framework for adaptive rigging and skinning of stylized character meshes using skeletal motion data. Our approach consists of two key stages: Skeletal Articulation Prediction, which adjusts the given skeleton to align with the source mesh, and Skinning Weight Prediction, which generates plausible skinning weights for the predicted target skeleton and given source mesh. One of the key strengths of our method is its flexibility to handle a wide variety of input formats, including variations in both skeletal and mesh configurations. By incorporating Diff3F as a semantic prior, our method can effectively model the correspondences between mesh vertices and skeleton joints, improving its generalizability to unseen characters. We demonstrated that our method outperforms previous rigging and skinning methods, along with the robustness of our method on rigging, skinning, and deformation given varying skeleton and mesh input.

\section*{Acknowledgements}
This work was supported by the National Research Foundation of Korea (NRF) grant funded by the Korea government (MSIT) (RS-2024-00333478).

\bibliographystyle{eg-alpha-doi} 
\bibliography{99.Ref}

\appendix
\section{Detailed explanation of $D^T$} \label{Appendix_A}
A skeletal motion data $M=({S, D^{1:N_T}})$, includes $S$ denoting the skeleton and $D^{1:N_T}$ representing the motion data with $N_T$ frames. Following Lee et al.~\cite{lee2023same}, $D^t$ is represented as follows:
\begin{align}
    D^t=\{ \mathbf{q}^t_{1:N_J}, \mathbf{p}^t_{1:N_J}, \mathbf{p}^{t-1}_{1:N_J}, \mathbf{v}^t_{1:N_J}, \mathbf{r}^t, \mathbf{c}^t_{1:N_J} \},
\end{align}
where $\mathbf{q}_j \in \mathbb{R}^6$ represents the local joint rotation with respect to the parent joint in the 6D representation~\cite{zhou2019continuity}. This representation is derived by taking the first two columns of a $3 \times 3$ rotation matrix, which ensures continuity and resolves ambiguities, stabilizing neural networks training. $\mathbf{p}_j \in \mathbb{R}^3$ represents joint positions in the character's facing frames~\cite{holden2017phase}. Specifically, a lateral vector is computed by averaging the vector between the hip joints and the vector between the shoulder joints, and then the facing direction is calculated as the cross product of the lateral and upward directions. Subsequently, the lateral vector is orthogonalized via the Gram-Schmidt process using the facing and upward directions, establishing a consistent 3D orthonormal basis. The origin of this frame is computed by projecting the root joint position onto the horizontal plane. The linear velocity $\mathbf{v}_j^t$ is computed by $\mathbf{v}_j=(\mathbf{p}_j^t-\mathbf{p}_j^{t-1}) / \Delta{t}$ as a translational difference between two consecutive frames. $\mathbf{r}^t=(\Delta{x}, \Delta{z}, \Delta{\theta}, h)$ denotes the root movement, where $(\Delta{x}, \Delta{z})$ and $\Delta{\theta}$ are translational velocities on the horizontal plane and rotational velocity around the up-axis with respect to the facing frame at the previous frame, while $h$ is the height of the root joint from the ground. Finally, $\mathbf{c}_j$ represents a contact label, indicating whether the $j$-th joint is in contact with the ground or not. The generated target pose at frame $t$ is defined as follows:
\begin{align}
    \hat{D}^t=\{ \mathbf{\hat{q}}^t_{1:N_J}, \mathbf{\hat{r}}^t, \mathbf{\hat{c}}^t_{1:N_J} \},
\end{align}
by omitting redundant elements such as joint positions, which can be derived by solving forward kinematics using $\mathbf{\hat{q}}$ and $\mathbf{\hat{r}}$.

\newpage
\section{Architecture Details}
Tables~\ref{tab:skeletal_articulation_prediction} and ~\ref{tab:skinning_weight_prediction} show the detailed network architectures of the Skeletal Articulation Prediction module and Skinning Weight Prediction module, respectively. The names correspond to those in Figure 2 of the main paper, except the Skeleton Decoder of Skeletal Articulation Prediction module in Table~\ref{tab:skeletal_articulation_prediction} refers to the combination of the Attention and MLP layers. 

\begin{table*}
\centering
\caption{Network architectures of the Skeletal Articulation Prediction module.}
\begin{tabular}{cccc}
\toprule
Name & Layers & Channels & Attention Heads \\
\midrule
\multirow{5}{*}{Mesh Encoder} & Linear - BatchNorm - ReLU - Dropout & 35 $\rightarrow$ 256 & - \\
                              & Linear - BatchNorm - ReLU - Dropout & 256 $\rightarrow$ 256 & - \\
                              & Linear - BatchNorm - ReLU - Dropout & 256 $\rightarrow$ 32 & - \\
                              & Pooling \& Concatenation & 32 $\rightarrow$ 64 & - \\
                              & Linear & 64 $\rightarrow$ 32 & - \\
\midrule
\multirow{4}{*}{Skeleton Encoder} & GAT - BatchNorm - ReLU - Dropout & 6 $\rightarrow$ 16 & 16 \\
                                  & GAT - BatchNorm - ReLU - Dropout & 256 $\rightarrow$ 16 & 16 \\
                                  & GAT - BatchNorm - ReLU - Dropout & 256 $\rightarrow$ 16 & 16 \\
                                  & GAT - BatchNorm - Dropout & 256 $\rightarrow$ 32 & 1 \\
\midrule
\multirow{4}{*}{Skeleton Decoder} & CrossAttention (QKV) - Dropout - Residual Connection & 32 $\rightarrow$ 2 & 16 \\
                                  & Linear - ReLU & 32 $\rightarrow$ 32 & - \\
                                  & Linear - ReLU & 32 $\rightarrow$ 32 & - \\
                                  & Linear - BatchNorm & 32 $\rightarrow$ 3 & - \\
\bottomrule
\end{tabular}
\label{tab:skeletal_articulation_prediction}
\end{table*}

\begin{table*}[t]
\centering
\caption{Network architectures of the Skinning Weight Prediction module.}
\begin{tabular}{cccc}
\toprule
Name & Layers & Channels & Attention Heads \\
\midrule
\multirow{4}{*}{Skeleton Encoder} & GAT - BatchNorm - ReLU - Dropout & 6 $\rightarrow$ 16 & 16 \\
                                  & GAT - BatchNorm - ReLU - Dropout & 256 $\rightarrow$ 16 & 16 \\
                                  & GAT - BatchNorm - ReLU - Dropout & 256 $\rightarrow$ 16 & 16 \\
                                  & GAT - BatchNorm - Dropout & 256 $\rightarrow$ 32 & 1 \\
\midrule
Skinnning Weight Predictor & CrossAttention (QK) & $N_V \times 32$ and $N_J \times 32$ $\rightarrow$ $N_V \times N_J$ & 1 \\
\bottomrule
\end{tabular}
\label{tab:skinning_weight_prediction}
\end{table*}

\section{Additional Experiments}
\subsection{Evaluations on Out-of-Domain Characters}
While our test dataset includes a wide range of stylized characters from Mixamo~\cite{mixamo}, we conducted additional experiments on out-of-domain characters to further evaluate the generalization capabilities of our method. Specifically, we employed characters in the T-pose from the RigNet-v1 dataset~\cite{xu2020rignet}, which are unseen during training. These characters were used as source meshes and deformed using source skeletons and motions from our database.

As shown in Figure~\ref{fig:OOD}, our method showed robust performance in generating target skeletons that align with the character mesh despite variations in the structures and shapes of the input skeletons. Furthermore, our method produced plausible skinning weights aligned seamlessly with both the source mesh and the predicted target skeletons, deriving plausible deformations for out-of-domain characters, which have shapes and body ratios distinct from Mixamo characters. These results highlight the robustness and generalizability of our approach on unseen characters. For animation results {with additional characters, please see the supplementary video.

\begin{figure}
    \centering    \includegraphics[width=\columnwidth]{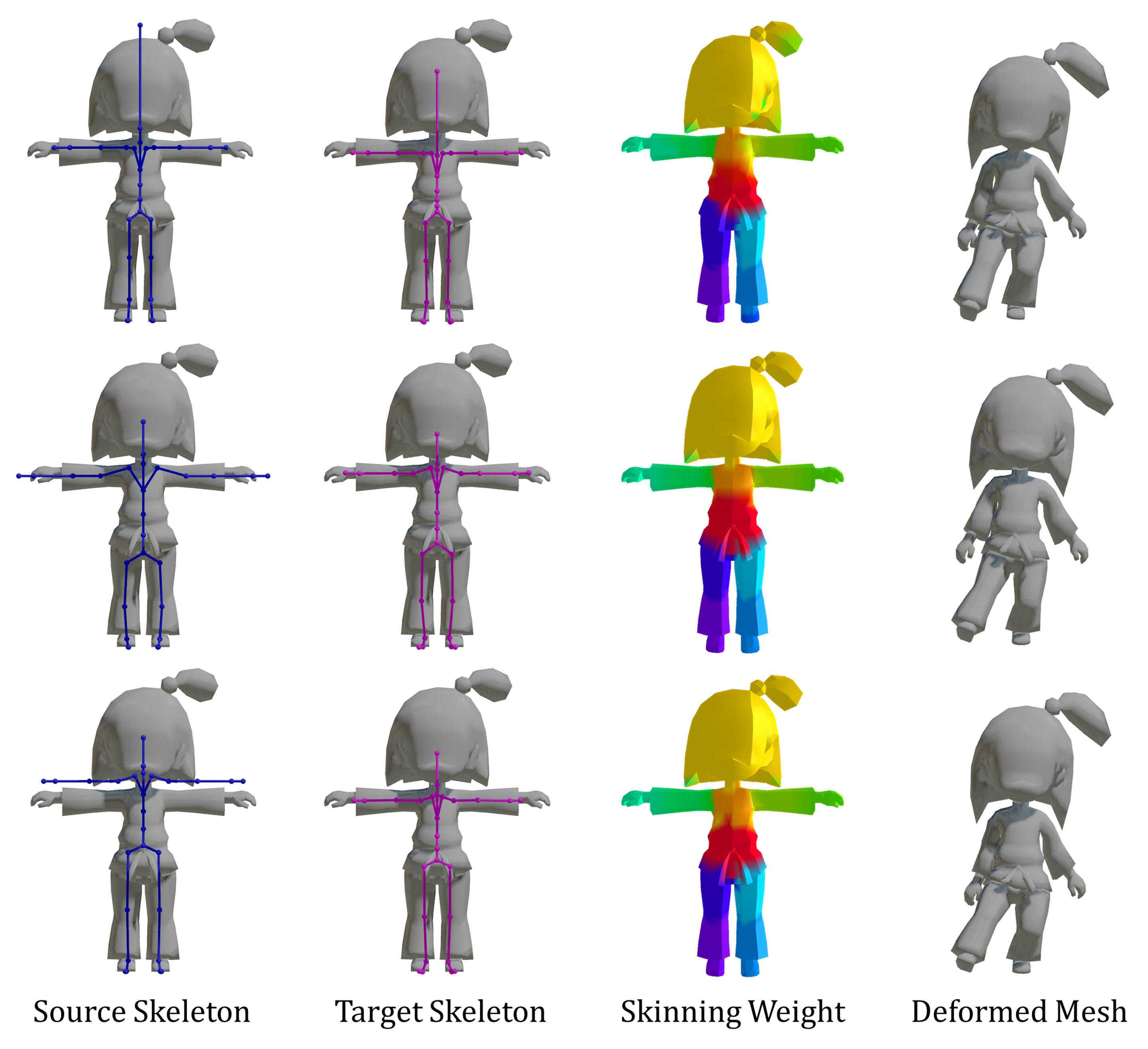}
    \caption{Qualitative results on out-of-domain characters from the RigNet-v1 dataset. Skeletons are overlaid on source meshes to demonstrate alignments between them.}
    \label{fig:OOD} 
    \vspace{-1em}
\end{figure}

\begin{table}
\centering
\caption{Quantitative results on skinning and deformation using identical source skeletons that precisely align with the source meshes. The best result for each column is in bold.}
\begin{tabular}{cccccc}
\hline
          & \makecell{Skinning \\ L1↓} & CD↓           & ADE↓          & MDE↓           & ELS↑          \\ \hline
Pinocchio & \textbf{0.0188}            & \textbf{2.83} & \textbf{2.56} & \textbf{13.74} & \textbf{0.96} \\ 
RigNet    & 0.0178                     & 2.95          & 4.71          & 95.60          & 0.10          \\ \hline
Ours      & 0.0469                     & 4.72          & 4.72          & 21.52          & 0.89          \\ \hline
\end{tabular}
\label{table:supp_skinning}
\vspace{-1em}
\end{table}

\subsection{Evaluations on Individual Impact} 
To rigorously evaluate the impact of individual changes in each independent variable, we conducted two additional experiments: (i) comparison of the performance of the skinning weight prediction module across different baselines, and (ii) analysis on the impacts of individual changes in each element of the skeletal configuration, including body scale, bone lengths, and the number of joints.

\noindent \textbf{Performance of Skinning Weight Prediction} \hspace{0.5mm} To solely compare the performance of the skinning weight prediction components across different baselines, we measured the skinning and deformation metrics using identical skeletons for all methods. Specifically, we used the source skeleton precisely aligned with the source mesh as input to the skinning weight prediction modules of each baseline to generate skinning weights, while bypassing their skeleton prediction modules. For deformation metrics, we used the source poses directly, instead of retargeting poses using SAME~\cite{lee2023same}. NBS~\cite{li2021nbs} was excluded from this experiment because its skinning weight prediction relies solely on the mesh and does not utilize skeleton inputs. To obtain the results of Pinocchio~\cite{baran2007automatic}, we followed the experimental setup of NBS~\cite{li2021nbs} that uses the auto-skinning tool provided by Blender, which is implemented based on the algorithm of Pinocchio.

As shown in Table~\ref{table:supp_skinning}, Pinocchio~\cite{baran2007automatic} achieved the best performance across all metrics, with our method producing comparable results. While ours did not achieve the best quantitative scores, the discrepancies were minor with visually imperceptible variations as shown in Figure~\ref{fig:skw_only}. Furthermore, our method still has an advantage in generalizability of rigging and skinning for various skeletal configurations, in that our method produced results consistent to Table 2 of the main paper, whereas Pinocchio produced results with significant deviation. RigNet~\cite{xu2020rignet} achieved comparable results in CD and ADE metrics to other methods, but its results exhibited noticeable artifacts, such as stretched vertices, which were reflected in significantly higher MDE and lower ELS values than those from other methods.

\begin{figure}
    \centering    \includegraphics[width=\columnwidth]{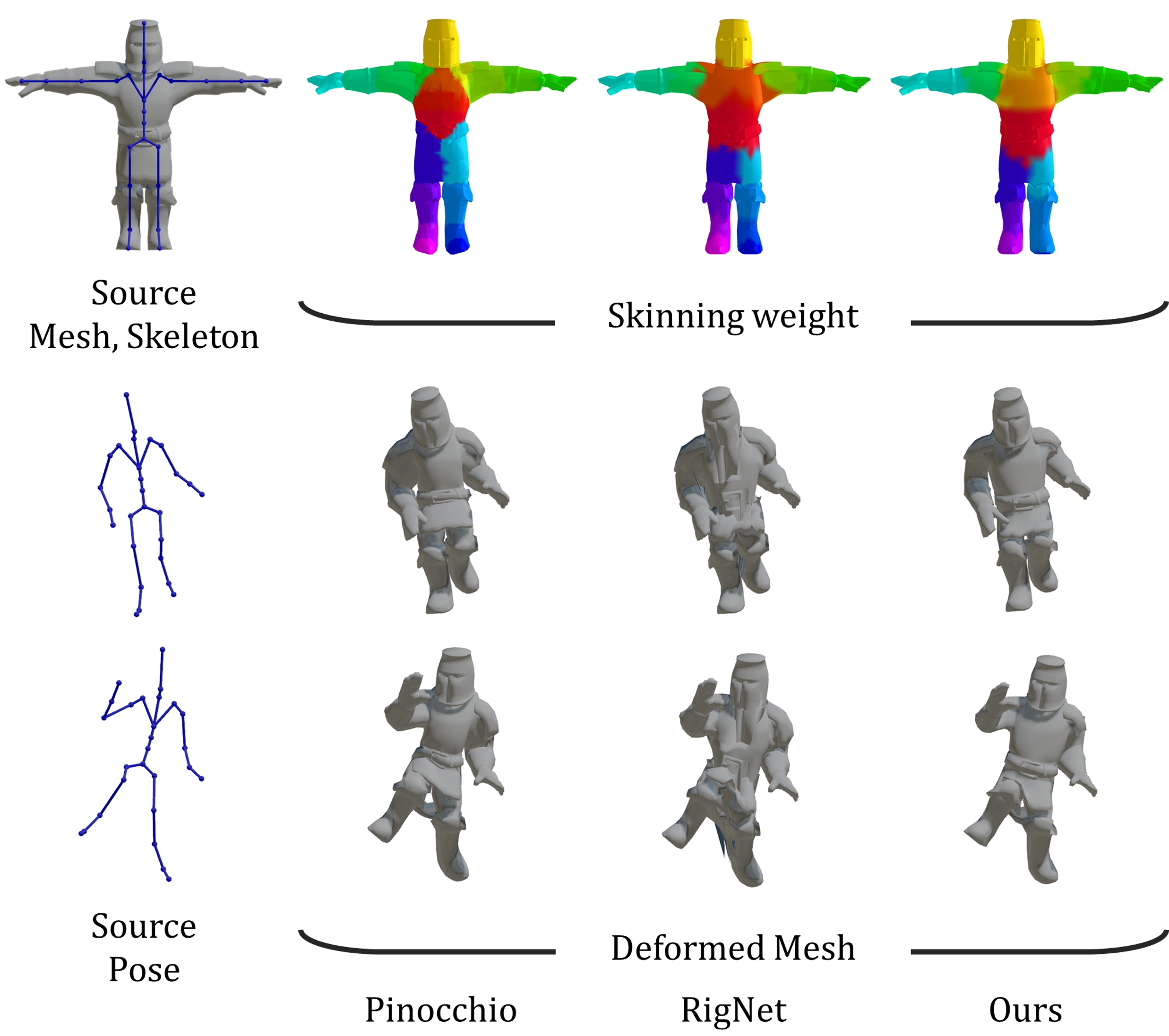}
    \caption{Qualitative comparison with baselines on predicted skinning weights and mesh deformation. The first column shows the source mesh and skeleton given to the skinning weight prediction module of each baseline, with source poses to deform the mesh. In the second to fourth columns, the first row demonstrates the skinning weights predicted by each baseline, while the second and third rows show the resulting deformed meshes.}
    \label{fig:skw_only} 
\end{figure}

\begin{figure*}[t]
    \centering    
    \includegraphics[width=\textwidth]{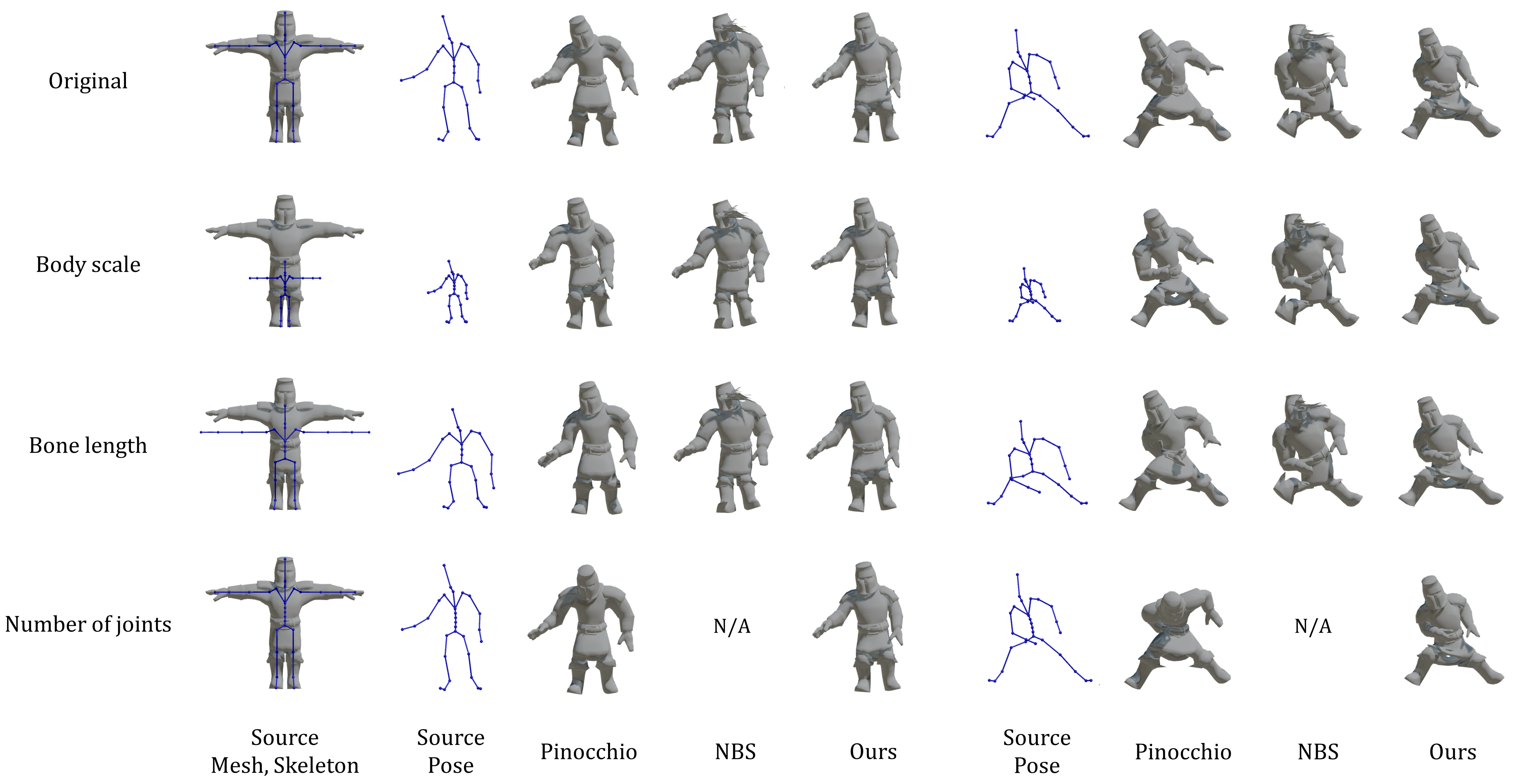}
    \caption{Qualitative comparison of deformation results under varying skeletal configurations. Beginning with a source skeleton that precisely aligns with the source mesh, each subsequent row includes the source skeleton generated by modifying one factor: body scale, bone length, or the number of joints, respectively, while maintaining the other elements fixed.}
    \label{fig:Separate_changes} 
\end{figure*}

\noindent \textbf{Changes of Individual Skeletal Configuration} \hspace{0.5mm} Starting with an initial source skeleton containing 25 joints, precisely aligned in size and proportion with the source mesh, we modified three key elements of the skeletal configuration to generate new source skeletons, as follows:
\begin{itemize}
  \item Body scale: A uniform scaling factor of 0.5 was applied to all bones to adjust the overall body size.
  \item Bone length: A non-uniform scaling was applied using scaling factors of 0.8 and 1.2 along the vertical and lateral axes of each joint, respectively.
  \item Number of joints: The number of joints was randomly adjusted, resulting in two additional joints to the initial skeleton. 
\end{itemize}
Figure~\ref{fig:Separate_changes} shows the generated source skeletons with their corresponding poses, along with the deformed meshes driven by each method based on the source mesh. Our approach consistently produced plausible deformations that follow the source poses, regardless of changes in the skeletal configuration. In contrast, Pinocchio~\cite{baran2007automatic} failed to preserve the volume of the source mesh, resulting in excessive expansion around shoulders and contraction around the spine. NBS~\cite{li2021nbs} resulted in distorted meshes due to improper skinning weights applied to certain body parts. Because NBS relies on a pre-defined set of joints, the results for the last source skeleton, which contains additional joints, were excluded.

\vfill\eject

\end{document}